\documentclass[aps,preprint,pre]{revtex4-1}

\usepackage[utf8]{inputenc}
\usepackage{float}
\usepackage[fleqn]{amsmath}
\usepackage[title]{appendix}
\usepackage{amsmath,geometry,amssymb,verbatim,graphicx,caption,makecell,subcaption,tikz,color,hyperref,float}
\usepackage{geometry}
\usepackage{enumitem}
\usepackage{siunitx}
\begin{document}

\title{Domain Growth in Ferronematics: Slaved Coarsening, Emergent Morphologies and Growth Laws}

\author{Aditya Vats}
\author{Varsha Banerjee}
\affiliation{Department of Physics, Indian Institute of Technology Delhi, New Delhi - 110016, India}
\author{Sanjay Puri}
\affiliation{School of Physical Sciences, Jawaharlal Nehru University, New Delhi - 110067, India}

\begin{abstract}
Ferronematics (FNs) are suspensions of magnetic nanoparticles in nematic liquid crystals (NLCs). They have attracted much experimental attention, and are of great interest both scientifically and technologically. There are very few theoretical studies of FNs, even in equilibrium. In this paper, we study the non-equilibrium phenomenon of domain growth after a thermal quench (or {\it coarsening}) in this coupled system. Our modeling is based on coupled {\it time-dependent Ginzburg-Landau} (TDGL) equations for two order parameters: the LC tensor order parameter ${\bf Q}$, and the magnetization ${\bf M}$. We consider both shallow and deep quenches from a high-temperature disordered phase. The system coarsens by the collision and annihilation of topological defects. We focus on {\it slaved coarsening}, where a disordered ${\bf Q}$ (or ${\bf M}$) field is driven to coarsen by an ordered ${\bf M}$ (or ${\bf Q}$) field.  We present detailed results for the morphologies and growth laws, which exhibit unusual features purely due to the magneto-nematic coupling. To the best of our knowledge, this is the first study of non-equilibrium phenomena in FNs. 
\end{abstract}

\maketitle

\section{Introduction}
\label{s1}

Nematic liquid crystals (NLCs) are a class of soft materials that combine fluidity of liquids with long-range orientational order of solids. The constituent nematic molecules can spontaneously align along a locally preferred direction, i.e., the {\it director}. This directionality results in physical properties such as anisotropic scattering and birefringence, and anisotropic response to external fields, incident radiation, temperature and colloidal inclusion. As a result, NLCs have acquired the status of smart materials, and are harnessed in many applications, the most significant being the LC display industry. Traditionally these applications have relied on the dielectric anisotropy or the direction-dependent response to electric fields. The anisotropy in the magnetic susceptibility is much smaller ($\sim 10^{-6}$); so large magnetic fields ($\sim 10$ G) are required for actuation \cite{DGen_1969}. Consequently, the magnetic effects in NLCs have not been utilized in applications. 

In 1970, Brochard and de Gennes in pioneering work proposed inclusion of a small amount of ferromagnetic particles in NLCs ($\sim 0.01\%$ by volume) to intensify magnetic sensitivity \cite{Broc_1970}. They predicted that the mechanical coupling due to surface anchoring of the magnetic grains by the surrounding nematic matrix could raise the susceptibility by several orders of magnitude in comparison to pure nematics. Although intense efforts were made to create these stable and highly sensitive suspensions, it was only in 2013 that Mertelj et al. obtained the first such using micron-sized disc-shaped barium hexaferrite (BaHF) magnetic nanoparticles (MNPs) in pentylcyano-biphenyl (5CB) NLCs \cite{Mert_Na2013}. Since then, this intriguing class of materials called {\it ferronematics} (FNs), is enjoying attention from academia and industry as well \cite{Mert_Soft2014, Zhng_PRL2015, Acke_Nmat2017, Mert_Ncomm2016, Mert_LCR2017, TPostisk_PRL2017, Mert_PRE2018, Qliu_PNAS2016, JSBTai_PNAS2018,Rupn_LC2015,Zarubin_2018,SHL_Klapp_2015,A_zakhlev_2016,Gaurav_2019,Gaurav_2020,Dierking_2020}. Clearly, the exploitation of the full range of their properties requires suitable experimental and theoretical study. 

For dilute MNP-NLC suspensions, it is reasonable to treat the nematic order and the averaged magnetic moment of the nanoparticles as continuous variables \cite{konark_2019,Konark_2019_2}. In this limit, a convenient theoretical description is based on the phenomenological Ginzburg-Landau (GL) free energies obtained by expansion in terms of appropriate order parameters \cite{JP_DG_1995}. These GL free energies are characterized by few constants that can be determined experimentally. FNs require two order parameters for their description: (i) the ${\bf Q}$-tensor which contains information about the orientational anisotropy of the LC; and (ii) the magnetization vector ${\bf M}$ which is the locally averaged magnetic moment of the suspended MNPs. The free energy of this complex fluid contains contributions from  magnetic and nematic counterparts, and in conjunction with a magneto-nematic coupling term. The latter plays a significant role in determining the static and dynamic properties of FNs. The minimization of the free energy yields different phases of the coupled system as a function of the thermodynamic parameters, e.g., temperature, pressure, etc.

An important non-equilibrium phenomenon is the {\it kinetics of phase transition} initiated by quenching a system from the disordered phase ($T>T_c$, where $T_c$ is the critical temperature) to the ordered phase ($T<T_c$). Such a system does not order instantaneously. Instead, there is formation of domains of degenerate ground states which are separated by interfaces or defects \cite{Puri_2009,Bray_2002}. The domains {\it coarsen} by eliminating the defects, and the system develops a characteristic length scale $L(t)$ which grows with time. The domain growth laws convey significant details of the ordering system, and the late stage growth kinetics can be explained in terms of the defect dynamics. Phase ordering studies of two dimensional ($d=2$) thermotropic NLCs were initiated in the early 1990s, and continue to remain interesting even today \cite{Blundell_1992,Puri_1993,Zapotocky_1995,Yeomans_2001,Bray_2002,Amit_2009,Amit_2010_2,Asingh_2012,Asingh_2014}. One of the reasons is that the LC order parameter ${\bf Q}$ is not a scalar or a vector as is usually the case, but a traceless symmetric tensor containing information about the orientational liquid crystalline anisotropy \cite{Mtram_ArX_2014}. Further, NLCs are experimental realizations of systems with continuous symmetry. They display stable topological defects with integer and half integer charges, which interact and annihilate in the coarsening process \cite{Zapotocky_1995,Puri_1993}. The late-stage coarsening controlled by this defect dynamics is believed to be distinct from that of $O(n)$ models with $n$-component vector order parameters. This is because the $O(n)$ models lack the inversion symmetry of the director field ${\bf n}$ \cite{Puri_1993,Bray_2002}. 

In a recent letter, we performed benchmarking studies of coarsening in $d=2$ FNs \cite{Aditya_2020}. We studied the kinetics of phase transitions via the time-dependent Ginzburg-Landau (TDGL) equations with a suitable GL free energy. Our observations in this letter are as follows:\\ 
(i) For shallow quenches ($T_c^N<T<T_c^M$, where $N$ and $M$ refer to the nematic and magnetic components), the ordering magnetic component can {\it enslave} the nematic component to coarsen. In that case, their domains are co-aligned. Similar statements hold for quenches such that $T_c^M<T<T_c^N$.  \\
(ii) For asymmetric coupling, there are sub-domain morphologies (SDM) dominated by interfacial defects;\\
(iii) The structure factor $S({k})$ for the SDM exhibits Porod decay, $S({k})\sim {k}^{-(d+1)}$, which is characteristic of scattering from sharp interfaces in dimensionality $d$. This contradicts our naive expectation of the generalized Porod tail $S({ k})\sim { k}^{-(d+2)}$ for scattering from vortex defects in continuous-spin models \cite{Bray_Puri_1991}.

This paper undertakes the challenging task of expanding our theoretical framework to $d=3$ FNs. The $d=3$ NLCs have a broader experimental relevance, and exhibit rich defect structures in the form of hedgehogs and strings. Surprisingly, there are very few studies of domain growth in $d=3$, even for pure NLCs \cite{Puri_1993,Nikolai_2002,Amit_2010}. This is probably because such a study involves the non-trivial task of obtaining the solutions of a set of 5 coupled nonlinear partial differential equations for the 5 independent components of ${\bf Q}$. On inclusion of MNPs, the system evolution is described by 8 coupled partial differential equations. In this paper, we undertake a comprehensive study of FNs in $d=2,3$. We elaborate the numerical techniques, provide detailed analytical results for $d=2$, and present detailed numerical results for $d=2,3$.

Our paper is organized as follows. In Section~\ref{s2}, we present the GL free energy models for FNs and formulate the corresponding TDGL equations. Then we study the fixed points and perform their linear stability analysis (in $d=2)$ to understand the role of the magneto-nematic coupling on coarsening in FNs. The tools required to characterize morphologies and derivations of growth laws are also discussed. Section~\ref{s3} provides detailed numerical results for $d=2,3$ FNs. The paper concludes with a summary and discussion of results in Section~\ref{s4}. The appendices contain details of calculations and tables which summarize results for the different quenches that we have considered.   

\section{Theoretical Framework}
\label{s2}

\subsection{Ginzburg-Landau (GL) Free Energy}
\label{s21}

The stability of FNs in the experiments of Mertelj et al. is believed to be a delicate consequence of the shape of the MNPs and the anchoring of the surrounding NLCs. A particle immersed in a liquid crystalline medium deforms the director field around it. This deformation can be expanded as a series of multipoles. In the absence of external torque on the particle, the monopole contribution is zero due to the uniaxial symmetry of the nematic molecules. The leading contributions in the expansion (dipolar, quadrupolar, etc.) depend on the shape of the particle and the nature of anchoring of the NLC molecules on the particle surface. Further the deformation introduces a {\it topological charge} $\pm1$ or $\pm1/2$. Dipolar interactions yield point defects, but quadrupolar interactions yield (two) disclination loops or a {\it Saturn ring} around the particle. Mertelj et al. used disc-like particles of BaHF with homeotropic anchoring. They argued that the director field around a single particle has quadrupolar symmetry, thereby introducing a topological charge $+1$ which needs to be compensated by the formation of other LC defects. For a quadrupole, the associated LC defect is a Saturn ring with a charge $-1$ \cite{Mert_LCR2017}. The stability of the FNs is believed to be a consequence of the repulsion between the LC defects of the same charge which dominates the dipole-dipole attraction between magnetic moments. 

{  There are two experimental protocols used to obtain stable FNs. In Method 1, the MNP-NLC mixture is quenched from the (disordered) isotropic phase to the (ordered) nematic phase in the presence of an external magnetic field applied along the director direction. This procedure results in a FN domain with {\it aligned} magnetic moments \cite{Mert_LCR2017}. Such suspensions are referred to as {\it ferromagnetic} NLCs. In this case, it is essential to incorporate an additional field-dependent term in the the free energy to represent the coupling between the applied field and the order parameter $\boldsymbol{M}$. In Method 2, the MNP-NLC suspension is quenched in the absence of an external magnetic field. The moments align along $\bf{n}$ or -$\bf{n}$ due to the magneto-nematic coupling and yield a vanishing macroscopic magnetization \cite{D_Petrov_2012,A_zakhlev_2016}.}

We use the Landau-de Gennes (LdG) approach which provides a phenomenological free energy in terms of the order parameters ${\bf Q}$ and ${\bf M}$. The ${\bf Q}$-tensor is symmetric and traceless, and is defined as: 
\begin{equation}
\label{QT}
Q_{ij}= S\left(n_i n_j-\frac{\delta_{ij}}{d}\right),
\end{equation}
where the amplitude $S$ is the scalar order parameter which measures the degree of order about the leading eigenvector or the director ${\bf n}$ \cite{AMaj_QT2012,konark_2019,Konark_2019_2}.  It can be approximated by the second order Legendre polynomial, $S=\overline{P_2(\cos{\theta})}$, where $\theta$ is the angle between nematic molecule and the director, and the over-bar indicates an average over all the molecules. The isotropic phase corresponds to $S=0$ and a fully aligned nematic phase has $S=1$. A nematic defect corresponds to regions of low order or $S \simeq 0$. The order parameter ${\bf M}$ measures the local magnetization of the MNPs. The typical GL free energy for the FN has the form \cite{JP_DG_1995, HPlein_2001, konark_2019,Konark_2019_2}:
\begin{equation}
\label{LdG_FE}
G(\boldsymbol{Q},\boldsymbol{M}) =\int \mbox{d}\bf{r}\left[g(\boldsymbol{Q},\boldsymbol{M}) + \frac{L}{2} \left|\nabla{\bf Q}\right|^2 + \frac{\kappa}{2}\left|\nabla \boldsymbol{M}\right|^2\right],
\end{equation}
where $g(\boldsymbol{Q},\boldsymbol{M}) $ is the local free energy density, and the two gradient terms account for the surface tension due to inhomogeneities in the order parameters. 

We use the free energy density proposed by Mertelj et al. \cite{Mert_Na2013} to obtain:
\begin{eqnarray}
\label{F_e}
G(\boldsymbol{Q},\boldsymbol{M})&=& \int \mbox{d}\bf{r} \left[ \frac{A}{2}\mbox{Tr}(\boldsymbol{Q}^2)+\frac{C}{3}\mbox{Tr}(\boldsymbol{Q}^3)+\frac{B}{4}\mbox{Tr}(\boldsymbol{Q}^2)^2+\frac{L}{2} \left|\nabla{\bf Q}\right|^2 \right.\nonumber\\ 
&& +\left.\frac{\alpha}{2}\left|\boldsymbol{M}\right|^2+\frac{\beta}{4}\left|\boldsymbol{M}\right|^4 + \frac{\kappa}{2}\left|\nabla \boldsymbol{M}\right|^2 -\frac{\gamma \mu_0}{2} \sum_{i,j=1}^{3}Q_{ij}M_iM_j\right].
\end{eqnarray}
Strictly speaking, the GL formulation is only valid near the critical point where the order parameter magnitudes are small. However, it has also been extensively applied to study systems far from criticality. The first four terms in Eq.~(\ref{F_e}) represent the GL free energy for the nematic component with Landau coefficients $A$, $B$, $C$, $L$ having their usual meaning. The next three terms correspond to the GL free energy for the magnetic component. The quadrupolar interactions have not been incorporated explicitly. Instead, the gradient term ${\vert\nabla {\bf M}\vert}^2$ captures the effects of magnetic interactions beyond a mean-field treatment by including short-range variations of the magnetic order parameter. The nematic free energy density in Eq.~(\ref{F_e}) includes terms up to Tr$(\boldsymbol{Q}^2)^2$ which yield the {\it uniaxial phase} in $d=3$. The {\it biaxial phase} in $d=3$ arises only when terms of order Tr$(\boldsymbol{Q}^6)$ or higher are included \cite{Luckhurst_2010}. The magneto-nematic coupling term is taken as a dyadic product of the ${\boldsymbol Q}$-tensor and ${\boldsymbol M}$ as it depends only on relative orientations, and respects the rotational invariance of the free energy.

We do not include any magnetic field term in our formulation as it introduces a directional bias, and may dominate the effects of the magneto-nematic coupling. Further, the stray field energy has also not been included in Eq.~(\ref{F_e}) as it is 1-2 orders of magnitude smaller than the nematic elastic energy contribution \cite{Mert_LCR2017,Hubert_2008}. Our study therefore bears relevance to FNs obtained via Method 2.

In Eq.~(\ref{F_e}), the Landau coefficients $ A = A_0(T-T_{c}^N)$ and $ \alpha = \alpha_{0}(T-T_{c}^M)$, where $A_0$ and $\alpha_0$ are positive constants. The Tr$(\boldsymbol{Q}^3)$ term with coefficient $C$ is relevant only in $d=3$, and the coefficient $C$ can be positive or negative. The parameters $B$ and $\beta$ are positive material-dependent constants, $L$ and $\kappa$ are the elastic constants, and $\gamma$ is the magneto-nematic coupling strength. These phenomenological parameters can be estimated from experimentally measured quantities. For example, $A$, $C$, $B$ and $L$ are related to the critical temperature, the latent heat of transition and the magnitude of the order parameter \cite{Priestly_2012}. Similarly, coefficients $\alpha$, $\beta$ and $\kappa$ can be evaluated from measurements of magnetization and susceptibility \cite{Hberg_2015}. The coupling constant $\gamma$ has been estimated by Mertelj et al. from the reversal fields of hysteresis loops  \cite{Mert_LCR2017}. However, the existing experimental data is for ferromagnetic NLCs prepared by Method 1, so numerical values of our model parameters are not available at this juncture.

\subsection{Time-dependent Ginzburg-Landau (TDGL) equations }
\label{s22}

We next discuss the methodology to study domain growth in FNs rendered thermodynamically unstable after a temperature quench. We provide a derivation of the time-dependent Ginzburg-Landau (TDGL) equations in dimensionless form for the $d=2$ FN. The corresponding calculations for $d=3$ are presented in Appendix \ref{app2}. While it is possible to calculate analytically the fixed points (FPs) for the $d=2$ case, this exercise is difficult for $d=3$ due to the large number of variables.

In $d=2$, the symmetric and traceless ${\bf Q}$-tensor has two independent components \cite{AMaj_QT2012,konark_2019,Konark_2019_2}:  
\begin{eqnarray}
\label{Q2}
{\bf Q}=
  \begin{pmatrix}
 Q_{11} &  Q_{12}   \\
 Q_{12} &  -Q_{11}   \\
  \end{pmatrix}.
\end{eqnarray}
{  It is easy to verify that $\mbox{Tr}({\bf Q}^2) = 2{\mid{\bf Q}\mid}^2= 2(Q_{11}^2+Q_{12}^2) = S^2/2$ and  $\mbox{Tr} ({\bf Q}^3) = 0$. Thus, we get a {\it continuous} nematic-isotropic transition in $d=2$. However, we consider quenches well below the nematic-isotropic transition temperature, so we do not expect consequences on the results due to the absence of the cubic term.}

{  The dissipative dynamics of FNs can be studied using coupled TDGL equations for the non-conserved order parameters $\boldsymbol{Q}$ and $\boldsymbol{M}$ \cite{Puri_2009,Bray_2002}:
\begin{eqnarray}
\frac{\partial \mathbf{Q}({\mathbf r},t)}{\partial t}&=&-\Gamma_Q\frac{\delta G[\mathbf{Q,M}]}{\delta \mathbf{Q}}\mbox{,} \label{TDGL_genral_1} \\
\frac{\partial \mathbf{M}({\mathbf r},t)}{\partial t}&=&-\Gamma_M\frac{\delta G[\mathbf{Q,M}]}{\delta \mathbf{M}},\label{TDGL_genral_2}
\end{eqnarray}
where the terms on the right are the functional derivatives of the free energy functional $G(\mathbf{Q,M})$ \cite{Puri_2009}. The coefficients $\Gamma_Q$ and $\Gamma_M$ are the damping factors for the nematic and magnetic components respectively. Using Eq.~(\ref{F_e}) with $C=0$, the order parameter evolution is given by:
\begin{eqnarray}
\label{tdgl1}
      \frac{1}{\Gamma_Q}\frac{\partial Q_{11}}{\partial t}&=&\pm 2|A|Q_{11}-4B|{\bf Q}|^2    Q_{11}+2L\nabla^2Q_{11}+\frac{\gamma \mu_0}{2}\left(M_1^2-M_2^2\right), \label{unscaled_TDGL_1}\\
\label{tdgl2}      
    \frac{1}{\Gamma_Q}\frac{\partial Q_{12}}{\partial t}&=&\pm2|A|Q_{12}-4B|{\bf Q}|^2   Q_{12}+2L\nabla^2Q_{12}+\gamma \mu_0\left(m_1m_2\right),\label{unscaled_TDGL_2}\\
\label{tdgl3}
    \frac{1}{\Gamma_M}\frac{\partial M_1}{\partial t}&=&\pm|\alpha|M_1-\beta|{\bf M}|^2M_1+\kappa\nabla^2M_1+\gamma \mu_0\left(Q_{11}M_1+Q_{12}M_2\right), \label{unscaled_TDGL_3}\\
\label{tdgl4} 
    \frac{1}{\Gamma_M}\frac{\partial M_2}{\partial t}&=&\pm|\alpha|M_2-\beta|{\bf M}|^2m_2+\kappa\nabla^2M_2+\gamma \mu_0\left(Q_{12}M_1-Q_{11}M_2\right).\label{unscaled_TDGL_4}
\end{eqnarray}
Notice that we consider the deterministic version of the TDGL equations and do not incorporate thermal fluctuations or noise. This is because noise is asymptotically irrelevant in domain growth problems \cite{Puri_1988}, except in the presence of quenched disorder \cite{Paul_2004,Paul_2005}.}
{  A dimensionless form of the TDGL equations can be obtained by introducing re-scaled variables $\mathbf{Q} = a\mathbf{Q^\prime}$, $\mathbf{M} = b\mathbf{M^\prime}$, $\textbf{r} = \xi\textbf{r}^\prime$,  $t=\tau t^\prime$. The appropriate choice for the scale factors is $a=\sqrt{|A|/2B}$, $b=\sqrt{|\alpha|/\beta}$, $\xi =\sqrt{\kappa/|\alpha|}$, $\tau = 1/(|\alpha| \Gamma_M$). Replacing these in Eqs.~(\ref{tdgl1})-(\ref{tdgl4}) and dropping the primes, we obtain the following TDGL equations for the FN in $d=2$ \cite{Aditya_2020}:
\begin{eqnarray}
\label{2TDGL1}
    \frac{1}{\Gamma}\frac{\partial Q_{11}}{\partial t}&=&\pm Q_{11}-|\mathbf{Q}|^2Q_{11}+l\nabla^2Q_{11}+c_1\left(M_1^2-M_2^2\right),\\
\label{2TDGL2}
    \frac{1}{\Gamma}\frac{\partial Q_{12}}{\partial t}&=&\pm Q_{12}-|\mathbf{Q}|^2Q_{12}+l\nabla^2Q_{12}+2c_1\left(M_1M_2\right),\\
\label{2TDGL3}
    \frac{\partial M_1}{\partial t}&=&\pm M_1-|\mathbf{M}|^2M_1+\nabla^2M_1+c_2\left(Q_{11}M_1+Q_{12}M_2\right),\\
\label{2TDGL4}
    \frac{\partial M_2}{\partial t}&=&\pm M_2-|\mathbf{M}|^2M_2+\nabla^2M_2+c_2\left(Q_{12}M_1-Q_{11}M_2\right).
\end{eqnarray}}
{  The dimensionless parameters in Eqs.~(\ref{2TDGL1})-(\ref{2TDGL4})  are}
\begin{equation}
\label{2dconstants}
    {  c_1=\frac{\gamma \mu_0 |\alpha|}{4|A|\beta}\sqrt{\frac{2B}{|A|}},
   \ \
 c_2=\frac{\gamma \mu_0 }{|\alpha|}\sqrt{\frac{|A|}{2B}},
\ \
     l=\frac{|\alpha|L}{2|A|\kappa},
\ \
    \Gamma= \frac{2|A|\Gamma_Q}{|\alpha|\Gamma_M}.}
\end{equation}

The $\pm$ sign with the first terms on the right depends on whether the order parameter $({\bf Q}$ or ${\bf M})$ is above ($-$) or below ($+$) its critical temperature. The parameters $c_1$ and $c_2$ are re-scaled coupling constants, $l$ sets the scale for relative diffusion of the nematic and magnetic components, and  $\Gamma$ is the relative damping coefficient. Note that $l$ affects only the non-universal prefactors of the growth laws and does not change the growth exponents, which are universal. The latter depend on (a) the dynamics of the system (conserved or non-conserved); (b) the nature of defects driving coarsening; and (c) the role of hydrodynamic effects \cite{Puri_2009,Bray_2002}. For simplicity, we set $l=1$. We also set $\Gamma=1$ subsequently. As $\Gamma$ is a rescaled parameter, the relaxation time-scales of ${\bf Q}$ and ${\bf M}$ can still be very different. Thus there remain only two phenomenological scaled constants in our simplified formulation: $c_1$ measuring the coupling strength of {$\bf Q$} and ${\bf M}$, and $c_2$ measuring the coupling strength of ${\bf M}$ with ${\bf Q}$. We emphasize that these originate from the same coupling term in Eq.~(\ref{F_e}). However, in our dimensionless re-scaling, they are combined with factors which determine the dimensional scales of the order parameters $\textbf{Q}$ and $\textbf{M}$ [see Eq.~(\ref{2dconstants})]. It is possible to determine our dimensional scales and the dimensionless coupling constants $c_1$ and $c_2$ from the Landau coefficients and the coupling constant $\gamma$ defined in the model free energy given by Eq.~(\ref{F_e}). This should be of relevance in experiments.

\begin{table}
\begin{center}
      \begin{tabular}{|l|l|}
       \hline
       Quench temperature & Coupling constants
        \\ 
       \hline
        \makecell{1) $T_c^N<T<T_c^M$~~~~~~ \\ 2) $T_c^M<T<T_c^N$~~~~~~ \\ 3) $T<\mbox{min}\{T_c^N,T_c^M\}$} & \makecell{(\romannumeral 1) $c_1\ne0$,\ $ c_2=0$ \\(\romannumeral 2)\ $c_1 = 0$,\ $ c_2\ne0$\\(\romannumeral 3) $c_1=c_2=c$~~~~} \\
        \hline
    \end{tabular}
    \end{center}
    \caption{Coarsening studies which we have undertaken.}
    \label{Tab1}
\end{table}
There are three cases of potential interest in this problem: (1) $T_c^N<T<T_c^M$, (2) $T_c^M<T<T_c^N$, and  (3) $T<\mbox{min}\{T_c^N,T_c^M\}$. We study these for the following sub-cases below: (\romannumeral 1) $c_1\ne0$, $c_2=0$, (\romannumeral 2) $c_1 = 0$, $c_2\ne0$, and (\romannumeral 3) $c_1=c_2=c$, as specified in Table \ref{Tab1}. A few remarks about the limiting cases are in order. An asymmetric coupling is not unusual, as the order parameters can have vastly different magnitudes in experimental systems, e.g., large magnetic particles in a bath of small LC molecules. Of course, we do not expect to realize $c_2=0$ or $c_1=0$, or $c_1=c_2=c$ precisely in experiments. However, it is possible that $c_1 \gg c_2$ or $c_2 \gg c_1$. We mimic these cases by (i)-(ii) above. Due to their analytical tractability, we shall see that the limiting cases provide useful guidelines for theoretical studies. A more quantitative discussion of the relevance of these limits for experimental systems will be presented shortly.

The scaled Eqs.~(\ref{2TDGL1})-(\ref{2TDGL4})  govern the evolution of the order parameters {\bf Q} and {\bf M} in $d=2$. (As mentioned earlier, the corresponding equations for $d=3$ are given in Appendix \ref{app2}.) Although our primary interest is in coarsening, it is useful to first study the long-time asymptotic limit  or the stationary solutions $({\bf Q^*,M^*})$ of these equations. These fixed points (FPs) determine the nature of domains which are formed in the process of coarsening. As there is no spatial or temporal variation in this limit, they can be obtained by setting $\partial \psi/\partial t$ and $\nabla^2\psi$ to 0 in Eqs.~(\ref{2TDGL1})-(\ref{2TDGL4}), where $\psi$ refers to the order parameters. This yields:
\begin{eqnarray}
\label{ss_case1}
   & &\pm {Q^*}_{11}-|\mathbf{Q^*|^2}{Q^*}_{11}+c_1({{M^*}_1}^2-{{M^*}_2}^2)=0,\label{st1}\\
  & &\pm  {Q^*}_{12}-|\mathbf{{Q^*}|^2}{Q^*}_{12}+2c_1({M^*}_1M^*_2)=0,\label{st2}\\
  & &\pm M^*_1-|{\bf M^*}|^2M^*_1+c_2(Q^*_{11}M^*_1+Q^*_{12}M^*_2)=0, \label{st3}\\
  & &\pm M^*_2-|{\bf M^*}|^2M^*_2+c_2(Q^*_{12}M^*_1-Q^*_{11}M^*_2)=0.\label{st4}
\end{eqnarray}

A trivial FP for Eqs.~(\ref{st1})-(\ref{st2}) is ${Q^*}_{11} = {Q^*}_{12} = {M^*}_{1} = {M^*}_{2} =0$. This corresponds to the high temperature disordered state. The explicit form of the non-trivial fixed points is given in Appendix \ref{app1} for cases 1 - 3 of Table \ref{Tab1}. We can also obtain the FPs analytically for arbitrary $c_1$, $c_2$, but these expressions are cumbersome and not presented here. To determine their stability, let us consider the evolution of small fluctuations around the stationary solutions ( ${\bf Q}^*+\Delta {\bf Q},\ {\bf M}^*+\Delta {\bf M} $), using the TDGL equations (\ref{2TDGL1})-(\ref{2TDGL4}). It is convenient to work with the Fourier transforms $\left[\Delta {\bf Q}({\bf k},t),\ \Delta {\bf M}({\bf k},t)\right]$. The corresponding equations, in the linear approximation, are given by:
\begin{eqnarray}\label{LS_equation}
 \frac{\partial \Delta Q_{11}}{\partial t}&=&2c_1M_1^*\Delta M_1 -2c_1M_2^*\Delta M_{2} +\left(\pm1-3{Q_{11}^*}^2-{Q_{12}^*}^2-{\bf k}^2\right)\Delta Q_{11} \nonumber \\
 && - 2Q_{11}^*Q_{12}^*\Delta Q_{12},  \label{LS_e1}\\
  \frac{\partial \Delta Q_{12}}{\partial t}&=&2c_1M_2^*\Delta M_1 +2c_1M_1^*\Delta M_{2}-2Q_{11}^*Q_{12}^*\Delta Q_{11}\nonumber\\ 
  && +\left(\pm1-3{Q_{12}^*}^2-{Q_{11}^*}^2-{\bf k}^2\right)\Delta Q_{12}, \label{LS_e2} \\
  \frac{\partial \Delta M_1}{\partial t}&=& \left(\pm1-3{M_1^*}^2-{M_2^*}^2+c_2Q_{11}^*-{\bf k}^2\right) \Delta M_1 +\left(c_2Q_{12}^*- 2M_1^* M_2^*\right) \Delta M_2 \nonumber\\
& & +c_2M_1^*\Delta Q_{11}+c_2M_2^*\Delta Q_{12},  \label{LS_e3}\\
\frac{\partial \Delta M_2}{\partial t}&=&\left(c_2Q_{12}^*- 2M_1^* M_2^*\right) \Delta M_1 + \left(\pm1-3{M_2^*}^2-{M_1^*}^2-c_2Q_{11}^*-{\bf k}^2\right) \Delta M_2 \nonumber \\
 & &  -c_2M_2^*\Delta Q_{11}+c_2M_1^*\Delta Q_{12}.  \label{LS_e4}
 \end{eqnarray}
Naturally, a solution is stable (unstable) if the fluctuations decrease (increase) with time.
Let us first consider the stability of the disordered solution with ${Q^*}_{11} = {Q^*}_{12} = {M^*}_{1} = {M^*}_{2} =0$. The stability properties are the same as the uncoupled case $c_1=c_2=0$. This is because the coupling terms do not contribute to the leading order. Thus growth in the ${\bf M}$-field (for $T<T_c^M$) cannot destabilize the ${\bf Q}$-field (for $T>T_c^N$) and vice-versa. Slaved coarsening is not possible in the linearized equations. As we shall see later, the fully non-linear equations admit the possibility of slaved coarsening.

Next, it is instructive to examine the solutions of Eqs.~(\ref{st1})-(\ref{st4}) and  Eqs.~(\ref{LS_e1})-(\ref{LS_e4}) for a typical case. We provide below an evaluation for Case 2(\romannumeral 2) where $T_c^M<T<T_c^N$ and $c_1=0,c_2\ne0$. Then, Eqs.~(\ref{st1})-(\ref{st4}) take the form: 
\begin{eqnarray}
    & & Q_{11}^*-|{\bf Q}^*|^2Q_{11}^*=0, \label{SS1_case22}\\
    & &Q_{12}^*-|{\bf Q}^*|^2Q_{12}^*=0,\label{SS2_case22}\\
    & &- M_1^*-|{\bf M}^*|^2M_1^*+c_2\left(Q_{11}^*M_1^*+Q_{12}^*M_2^*\right)=0,\label{SS3_case22}\\
    & &- M_2^*-|{\bf M}^*|^2M_2^*+c_2\left(Q_{12}^*M_1^*-Q_{11}^*M_2^*\right)=0.\label{SS4_case22}
\end{eqnarray}
If $\theta$ is the angle between ${\bf M}^*$ and $x-$axis, the non-trivial solution of Eqs.~(\ref{SS1_case22})-(\ref{SS4_case22}) is given by : 
\begin{align}\label{SS_case22}
  Q_{11}^*= \cos{2\theta}, \quad Q_{12}^*= \sin{2\theta}; \quad M_1^*=r_M\cos{\theta}, \quad M_2^*=r_M\sin{\theta};\end{align} 
with $\theta$ arbitrary and $ r_M =  \sqrt{c_2-1}$. It is easy to check using Eq.~(\ref{QT}), that the director ${\bf n}$ also makes an angle $\theta$ with the $x$-axis. Thus ${\bf n}$ and ${\bf M}$ are co-aligned in the stationary state, as is intuitively expected. This feature is ubiquitous to all the cases, and we will see that it has important consequences for the non-equilibrium properties of FNs. 

Next, we determine the stability of the stationary solution in Eq.~(\ref{SS_case22}). Given the rotational invariance, we can choose $\theta =0$ without loss of generality. The fluctuations in the Fourier component, from Eqs.~(\ref{LS_e1})-(\ref{LS_e4}) for $c_1=0$ and $c_2\ne0$, are as follows:
\begin{eqnarray}\label{LS_case22}
    \Delta Q_{11}({\bf k},t)&=&\Delta Q_{11}({\bf k},0) e^{-(2+{\bf k}^2)t}, \label{LS_Case22_1} \\
    \Delta Q_{12}({\bf k},t)&=&\Delta Q_{12}({\bf k},0) e^{-{\bf k}^2t},\label{LS_Case22_2}\\  
    \Delta M_1({\bf k},t)&=& \left[\Delta M_1({\bf k},0) - {c_2^\prime\Delta Q_{11}({\bf k},0)}\right] e^{(2-2c_2-{\bf k}^2)t} + {c_2^\prime\Delta Q_{11}({\bf k},0)}e^{-(2+{\bf k}^2)t},\label{LS_Case22_3}\\
    \Delta M_2({\bf k},t)&=& \left[ \Delta M_2({\bf k},0) - {c_2^{\prime\prime}\Delta Q_{12}({\bf k},0)}\right] e^{-(2c_2+{\bf k}^2)t} + {c_2^{\prime\prime}\Delta Q_{12}({\bf k},0)} e^{-{\bf k}^2t},\label{LS_Case22_4} 
  \end{eqnarray}
where we have defined $c_2^\prime=(c_2\sqrt{c_2-1})/(2c_2-4)$ and $c_2^{\prime\prime}=(\sqrt{c_2-1})/2$.  It is evident that, for an arbitrary ${\bf k}$, the fluctuations in ${\bf Q}$ always decrease. However, $\Delta M_1$ is stable only if $c_2>1$. (Of course, from Eq.~(\ref{SS_case22}), the non-trivial FP exists only for $c_2>1$.) Thus the linear stability analysis provides limits on $c_1$ and $c_2$, and facilitates their choice in simulations. 

\subsection{Morphology Characterization}
\label{s23}

A useful tool to understand evolving morphologies is the spatial correlation function defined in terms of the order parameter field $\boldsymbol{\psi}({\bf r},t)$ as \cite{Puri_2009}:  
\begin{equation}
\label{cr_def}
C({\textbf{r}},t)=\frac{1}{V}\int \mbox{d}{\bf R} \left[ \langle \boldsymbol{\psi}({\bf R},t) \cdot \boldsymbol{\psi}({\bf R}+{\bf r},t) \rangle - \langle \boldsymbol{\psi}({\bf R},t) \rangle \cdot \langle \boldsymbol{\psi}({\bf R} + {\bf r},t) \rangle \right] ,
\end{equation}
where, $V$ is the system volume, and $\langle \cdots \rangle $ indicates an averaging over independent runs. In Eq.~(\ref{cr_def}), we have assumed that the system is translationally invariant. We consider the case of isotropic systems, which are  characterized by a single length scale $L(t)$. The correlation function in such systems exhibits a dynamical scaling form \cite{Puri_2009}:
\begin{equation}
\label{cr_sc}
C({\textbf{r}},t) = f\left(\frac{ r}{L}\right),
\end{equation}
where $f(x)$ is the scaling function. The characteristic length scale $L(t)$ is defined as the distance over which the correlation function decays to (say) 0.5 of its maximum value. Small-angle scattering experiments yield the structure factor, which is the Fourier transform of the correlation function:
\begin{equation}
    S({\bf k},t) = \int d{\bf r}\ e^{\iota {\bf k}.{\bf r}}\ C({\bf r},t).
\end{equation}
The corresponding dynamical scaling form is given by
\begin{equation}
S({\bf k}, t)=L^{d}\ g({\bf k} L), 
\end{equation}
where $g(p)$ is the Fourier transform of $f(x)$. The characteristic length can also be defined as the inverse of the first moment of the structure factor \cite{Puri_2009}:
\begin{equation} \label{length_struc}
    L(t)= \Bigg[\frac{\int { \bf k}\ S({\bf k},t)\ d{\bf k}}{\int S({\bf k},t)\ d{\bf k} } \Bigg]^{-1},
\end{equation}

An approximate form of the correlation function for a system described by an $n$-component order parameter with non-conserved kinetics has been obtained by Bray and Puri \cite{Bray_Puri_1991} and Toyoki \cite{Toyoki_1992} by studying the defect dynamics. The Bray-Puri-Toyoki (BPT) function is valid for $n\leq d$, so that topological defects are present in the system. It has the following analytical form:
\begin{equation}\label{BPT}
          C(r,t)=\frac{n \gamma}{2 \pi} \Bigg[B \Bigg( \frac{n+1}{2},\frac{1}{2}\Bigg)\Bigg]^2 F\Bigg(\frac{1}{2},\frac{1}{2};\frac{n+2}{2};\gamma^2\Bigg).
\end{equation}
In the above expression, the beta function $B(x,y)=\Gamma(x)\Gamma(y)/\Gamma(x+y)$;  $F(a,b;c;z)$ is a hyper-geometric function \cite{Gradshteyn_2014}; and $\gamma = \exp\left[-r^2/(2L^2)\right]$ with $L$ being the average defect length. BP also demonstrated that the corresponding scaling function $g(p)$ exhibits the following large-$p$ or tail behavior \cite{Porod_1982,Yono_puri_1988}: 
\begin{equation}
g(p) \sim p^{-(d+n)} \quad \mbox{for} \quad p \rightarrow \infty.
\end{equation}
This result is referred to as the {\it generalized Porod law} as it generalizes the well-known Porod law, $g(p)\sim p^{-(d+1)}$, which characterizes scattering off sharp interfaces \cite{Porod_1982,Yono_puri_1988}.   
The structure factor tail conveys the dominance of different kinds of topological defects in the system \cite{Bray_Puri_1991}. For $n=1$, the defects are interfaces, and the corresponding scattering function exhibits  the {\it Porod law}. For $n>1$, the different topological defects are vortices ($n=2,\ d=2$), strings ($n=2,\ d=3$), and monopoles or hedgehogs ($n=3,\ d=3$). So in $d=3$ for example, $g(p) \sim p^{-5}$ or $\sim p^{-6} $ depending on whether strings or monopoles dominate in the defect dynamics.

The determination of the domain growth law $[L(t)$ vs. $t]$ is an important aspect in coarsening experiments. It reveals details of the free-energy landscape and relaxation time scales in the system. For example, pure isotropic systems with non-conserved dynamics obey the Lifshitz-Allen-Cahn law (LAC): $L(t) \sim t^{1/2}$ \cite{Allen_Cahn_1979}.  On the other hand, pure isotropic systems with conserved kinetics and diffusive transport follow the Lifshitz-Slyozov (LS) law: $L(t) \sim t^{1/3}$ \cite{Lif_Sly_1961}. These growth laws are characteristic of systems with no energy barriers to coarsening and a unique relaxation timescale. Let us understand how they arise -  this exercise will be useful to interpret the novel results in FNs presented shortly. The evolution of the order parameter $\psi$ for a non-conserved system with a scalar order parameter $\psi$ is governed by the TDGL equation: $\partial \psi/\partial t=\psi-\psi^3+\nabla^2\psi$, where we have used dimensionless units \cite{Puri_2009}. As $\psi$ has a {\it kink} profile at the interface between domains, the corresponding equation of motion can be obtained by rewriting the TDGL equation in terms of the interfacial coordinates \cite{Allen_Cahn_1979}. This yields the Allen-Cahn equation $\mbox{d}L/\mbox{d}t = -(d-1)/L$, which on integration results in the LAC law $L(t)\sim t^{1/2}$. BPT showed that the same growth law applies for the TDGL equation with vector order parameter, i.e., Eqs.~(\ref{2TDGL3})-(\ref{2TDGL4}) with $c_2 = 0$. However, when $n=2$, as in the $XY$ model in $d=2$, there may be logarithmic corrections to the growth law. 
 
For completeness, let us examine the growth law for a system with conserved order parameter, e.g., kinetics of phase separation in a binary (AB) mixture. In this case, a suitable order parameter is the density difference of the two species: $\psi = n_A -n_B$. For systems with only diffusive transport (solid mixtures), the evolution is described by the Cahn-Hillard (CH) equation:
\begin{equation}
\label{CHC}
\frac{\partial \psi}{\partial t} = \nabla^2\mu = \nabla^2[-\psi +\psi^3 - \nabla^2\psi],
\end{equation}
where again we have used dimensionless units \cite{Puri_2009}. As for the TDGL equation, we assume that the interface between domains is in local equilibrium. This allows us to obtain the conserved counterpart of the Allen-Cahn equation. This is considerably more complicated as the conservation constraint implies that the velocity at a point on the interface obeys an integral equation over all other interfaces of the system \cite{Bray_2002}. Nevertheless, this complicated equation is amenable to dimensional analysis. The chemical potential on the surface of a domain of size $L$  is $\mu \sim \sigma/L$, where $\sigma$ is the surface tension. The corresponding current is $j = \mid \nabla\mu\mid \sim \sigma/L^2$. The domain size grows as $\mbox{d}L/\mbox{d}t \sim \sigma/L^2$, which on integrating yields the LS law $L(t) \sim (\sigma t)^{1/3}$.

\section{Detailed Numerical Results}
\label{s3}

{  All simulations in $d=2$ FNs have been performed on a system of size $N^2=(2048)^2$, and the $d=3$ results have been obtained for $N^3=(256)^3$. Periodic boundary conditions are employed to remove the edge effects.  The initial values of the components of ${\bf Q}$ and ${\bf M}$ are small fluctuations about 0, to mimic the disordered system before the quench. The evolution is studied solving Eqs.~(\ref{2TDGL1})-(\ref{2TDGL4}) numerically using the Euler discretization method \cite{kin_Num2009}. The discretization mesh sizes $\Delta x$ and $\Delta t$ set the spatial and temporal scales for the system. The choices of these mesh sizes do not affect the nature of solutions as long as (a) they satisfy the stability criterion; and (b) the spatial mesh size $\Delta x$ is adequate to resolve the defect region \cite{Puri_1987,Puri_1988_PRA1,Puri_1988_PRA2,Elder_1988}. Recall the scalar TDGL equation $\partial \psi/\partial t = \psi - \psi^3 + \nabla^2\psi$. The corresponding stability condition is derived from the requirement that the Euler-discretized system should not suffer a sub-harmonic bifurcation about the stable FPs $\psi^* = \pm 1$. This yields the condition \cite{Puri_2011}:
\begin{equation}
\label{SC}
\Delta t < \frac{\Delta x^2}{\Delta x^2 + 2d}.
\end{equation}
The same condition applies for the vector TDGL equation, and we use it to determine the mesh sizes in our simulations. All numerical results presented are averaged over 10 independent runs (or more if required) to obtain clean numerics.

Before proceeding, it is useful to discuss the role of system size in the simulations. Our experience with domain growth problems \cite{Puri_1987,Puri_1988_PRA1,Puri_1988_PRA2} shows us that finite-size effects arise when the characteristic domain scale is larger than $\simeq 25$ \% of the lateral system size. These finite-size effects are signalled by a slowing-down of the domain growth law, and a consequent under-estimation of the growth exponent on a log-log scale. To eliminate artefacts due to the system size, we focus on results for time-windows where $L(t) < (N \Delta x)/4$.}

\subsection{Domain Growth in $d=2$ Ferronematics}
\label{s31}

{The $d=2$ simulations mimic experiments in which the length scale of the emergent morphologies is much larger than the thickness of the sample. Such a geometry has generally been realized in experiments by surface treatment of the top and bottom layers of the sample \cite{Tsakonas_2007}, and more recently via confinement between two substrates \cite{Nkumar_2018}.} {  We solved Eqs.~(\ref{2TDGL1})-(\ref{2TDGL4}) as described above on a square lattice of size $(N\Delta x)^2$. The mesh sizes are $\Delta x = 1$ and $\Delta t=0.01$.} The orientation of the director can be obtained from the $Q_{ij}$'s using Eq.~(\ref{QT}). The nematic and magnetic morphologies depict the orientation of ${\bf n}$ and ${\bf M}$ at each point on the square grid at the given time. The magnitudes agree with those of the stationary solutions, except at the defects where the respective order parameters are $\simeq 0$. We have obtained numerical results for all the cases in Table \ref{Tab1}, and present some representative results below. The results for other cases are summarized in the Tables \ref{TabS1}-\ref{TabS3} of Appendix \ref{app1}. 

Fig.~\ref{f1} shows the nematic (left) and magnetic (right) morphologies at $t=10^3$. Figs.~\ref{f1}(a)-(b) are for a quench temperature $T<\mbox{min}\{T_c^N,T_c^M\}$ and $c_1=c_2=0$, i.e., the uncoupled system. Recall that {\bf n} has an inversion symmetry. Therefore in the nematic snapshots, blue corresponds to {\bf n} in the first (or third) quadrant while green  corresponds to {\bf n} in the second (or fourth) quadrant. In the magnetization 
\begin{figure}[t]
\centering                      
\includegraphics[width=0.6\linewidth]{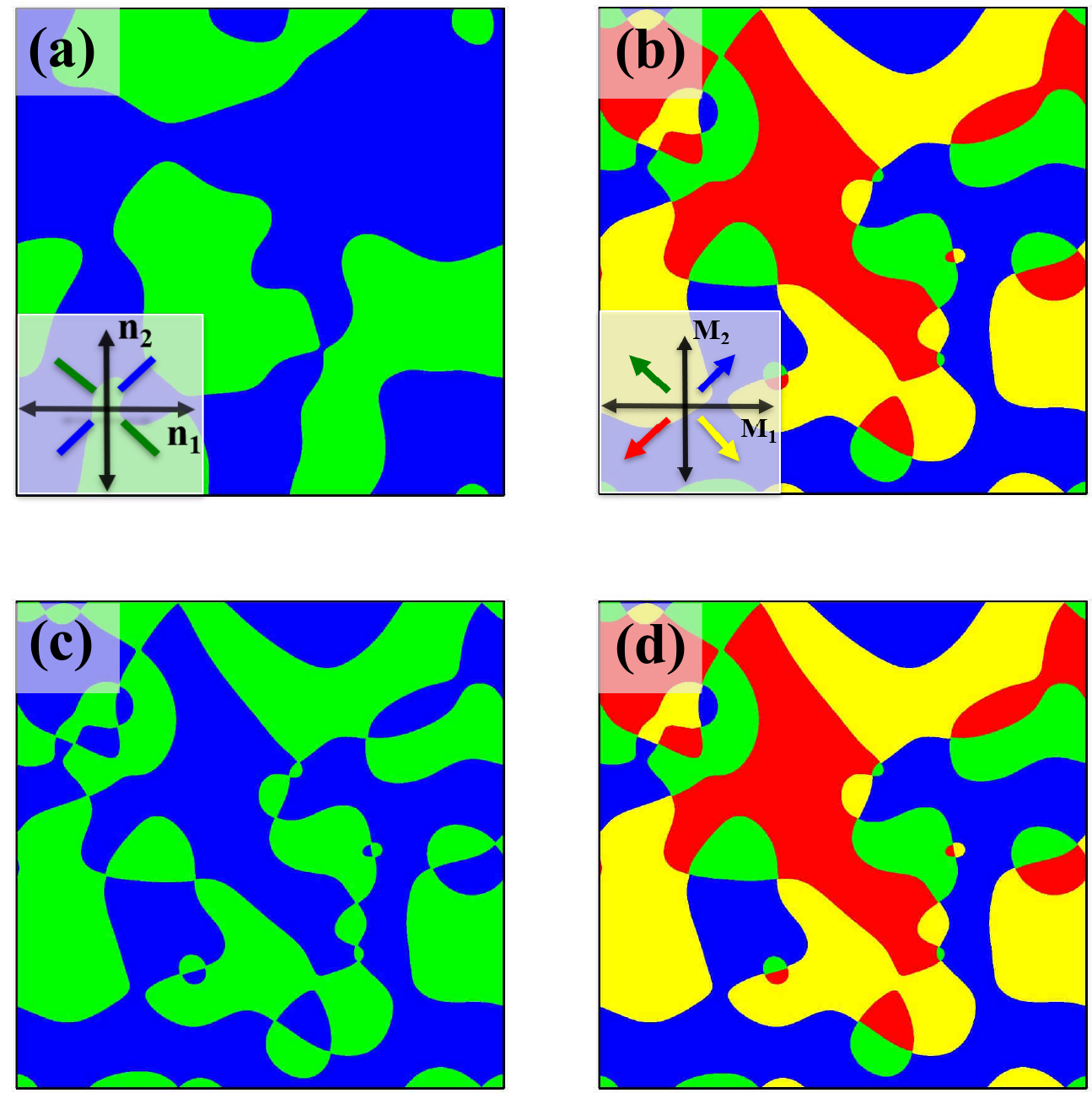} 
\caption{(color online) Domain growth in $d=2$ ferronematics: nematic morphologies (snapshots on left) and magnetic morphologies (snapshots on right) at $t=10^3$. The frames (a)-(b) correspond to the uncoupled case, i.e., $c_1=c_2=0$ with quench temperature $T<\text{min}\{T_c^N,T_c^M\}$. The frames (c)-(d) correspond to Case 1(\romannumeral 1), i.e., $c_1=4,\ c_2=0$ of Table \ref{Tab1}. The insets in (a)-(b) show the color scheme used to depict the order parameters.}
\label{f1}
\end{figure}
snapshots, four different colors depict $\textbf{M}$ lying in the first, second, third and fourth quadrants respectively. As expected, the two morphologies evolve independently due to the absence of the magneto-nematic coupling. Figs.~\ref{f1}(c)-(d) are for Case 1(\romannumeral 1) with $c_1=4,\ c_2=0$, and with the {\it same initial state} used in Figs.~\ref{f1}(a)-(b).  As $T_c^N<T<T_c^M$, linear stability analysis suggests an isotropic state for the nematic component, and  a ferromagnetic state for the magnetic component. The unusual feature here is {\it slaved coarsening} of the nematic phase due to coupling with the ordering magnetic component. The {\bf M}-field is linearly unstable and starts growing. When the amplitude has grown to a large magnitude, it drags the {\bf Q}-field into growth. We emphasize that this is a purely non-linear effect as our linearized equations are the same as those in the uncoupled case $c_1=c_2=0$. Notice that ${ \bf M}$ is always parallel (or anti-parallel) to ${ \bf n}$, as depicted in Figs.~\ref{f1}(c)-(d), due to the coupling term in the free energy in Eq.~(\ref{F_e}). Further, the magnitudes of ${\bf M}$ and ${\bf Q}$ are in accordance with the stationary solutions in Table \ref{TabS1} of Appendix \ref{app1}.

The spatial variation of {\bf n} and {\bf M} in Fig.~\ref{f1} and subsequent snapshots depends on the nature of defects in the relevant field. Thus, if the defects are interfaces (as we will see shortly for sub-domain morphologies or SDM), the domain walls are narrow. On the other hand, if the defects are vortices or strings, the order parameters vary smoothly.
\begin{figure}[t]
\centering
\includegraphics[width=0.7\linewidth]{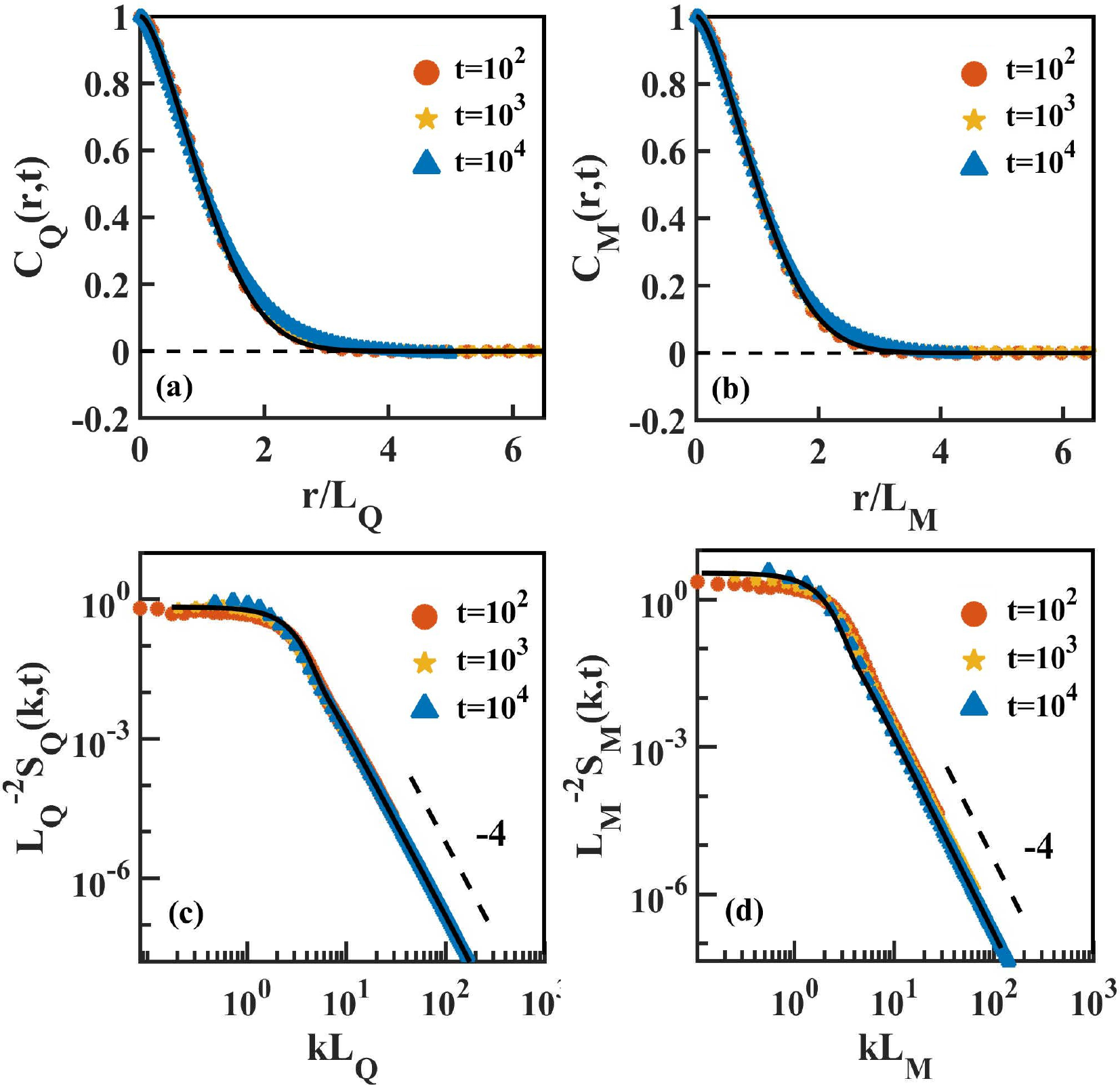} 
\caption{Dynamical scaling of correlation functions and structure factors in $d=2$. We show data for the nematic (left) and magnetic (right) components for Case 1(\romannumeral 1), i.e., $c_1=4,\ c_2=0$. The solid lines in the frames denote the BPT function [see Eq.~\ref{BPT}] evaluated for $n=2,d=2$.}
\label{f2}
\end{figure}

Next, we quantify characteristic features of the domains and defects by evaluating $C({r},t)$ and $S({k},t)$. The correlation function $C_M({r},t)$ is calculated directly from the definition in Eq.~(\ref{cr_def}) with ${\bf \psi}\rightarrow {\bf M}$. For the ${\bf Q}$-field, $C_Q({r},t)$ is calculated from the tensor order parameter ${\bf Q}$. There are two length scales in this problem, $L_Q$ and $L_M$, characterizing correlated regions in the nematic and magnetic components respectively. We define $L_Q$ $\left(L_M\right)$ to be the distance over which the correlation function decays to half its maximum value. Figs.~\ref{f2}(a)-(b) present data for $C_Q({r},t)\ \text{vs.}\ {r}/L_Q(t)$ and $C_M({r},t)\ \text{vs.}\ {r}/L_M(t)$ for Case 1(\romannumeral 1) with $c_1=4,\ c_2=0$ at $t=10^2,\ 10^3,\ 10^4$. Both sets of data exhibit an excellent collapse, indicating dynamical scaling. Thus, the morphologies of the nematic and magnetic domains in this coupled system do not change with time apart from a scale factor. The solid line in Figs.~\ref{f2}(a)-(b) is the BPT function calculated with $n=2,\ d=2$. These sets of data are well represented by the BPT function. In Figs.~\ref{f2}(c)-(d), we plot the corresponding structure factors $L_{Q}^{-2}S_Q(k,t)$ vs. $kL_Q$ and $L_{M}^{-2}S_M(k,t)$ vs. $kL_{M}$. The asymptotic structure factor tails of both components exhibit $k^{-4}$ behavior, indicating the dominance of vortex defects in ordering kinetics. In the generalized Porod tail, we set $d=2$, $n=2$. As a matter of fact, $S(k) \sim  k^{-4}$ is observed whenever (\romannumeral 1)$\ c_1 \ne 0,\ c_2=0$ and (\romannumeral 3)$\ c_1=c_2=c$, see Tables \ref{TabS1}-\ref{TabS3} in Appendix \ref{app1}.

As discussed in Sec.~II.B, we do not expect $c_1$ or $c_2$ to be precisely 0 in experimental systems. It is therefore useful to check the robustness of our results with respect to deviations from the limiting cases (i)-(iii) in Table~\ref{Tab1}. In Fig.~\ref{f3}, we plot the growth law $L(t)$ vs. $t$ (on a log-log scale) for nematic and magnetic components for Case 1 with $c_1=3$ and $c_2=0,1,2$. On the scale of this plot, the data for both components does not show deviations from the $c_2=0$ data even up to $c_2 \simeq 1$. Therefore, we expect that the results for limiting case (i) are valid for $c_1 \gg c_2$. Similar statements can be made for $c_2 \gg c_1$.
\setcounter{figure}{2}
\begin{figure}[t]
\centering                      
\includegraphics[width=0.4\linewidth]{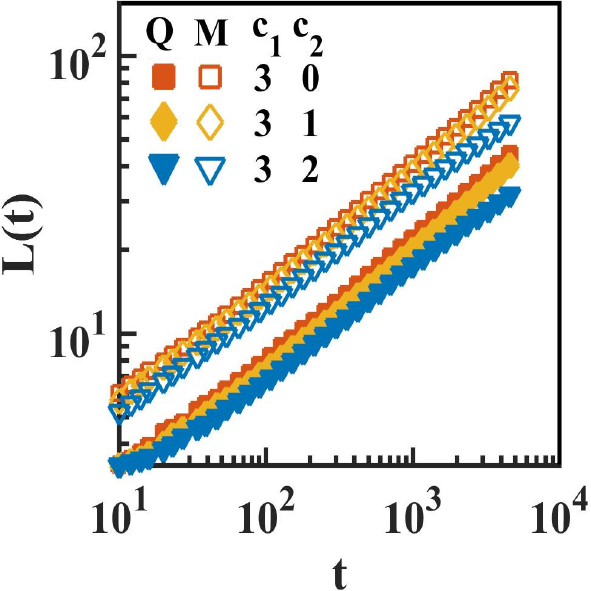} 
\caption{Growth laws in $d=2$ (on a log-log scale) for the nematic (solid symbols) and magnetic (open symbols) components for Case 1 in Table~\ref{Tab1} with $c_1=3$ and $c_2=0,1,2$.}
\label{f3}
\end{figure}

\begin{figure}[t]
\centering                      
\includegraphics[width=0.9\linewidth]{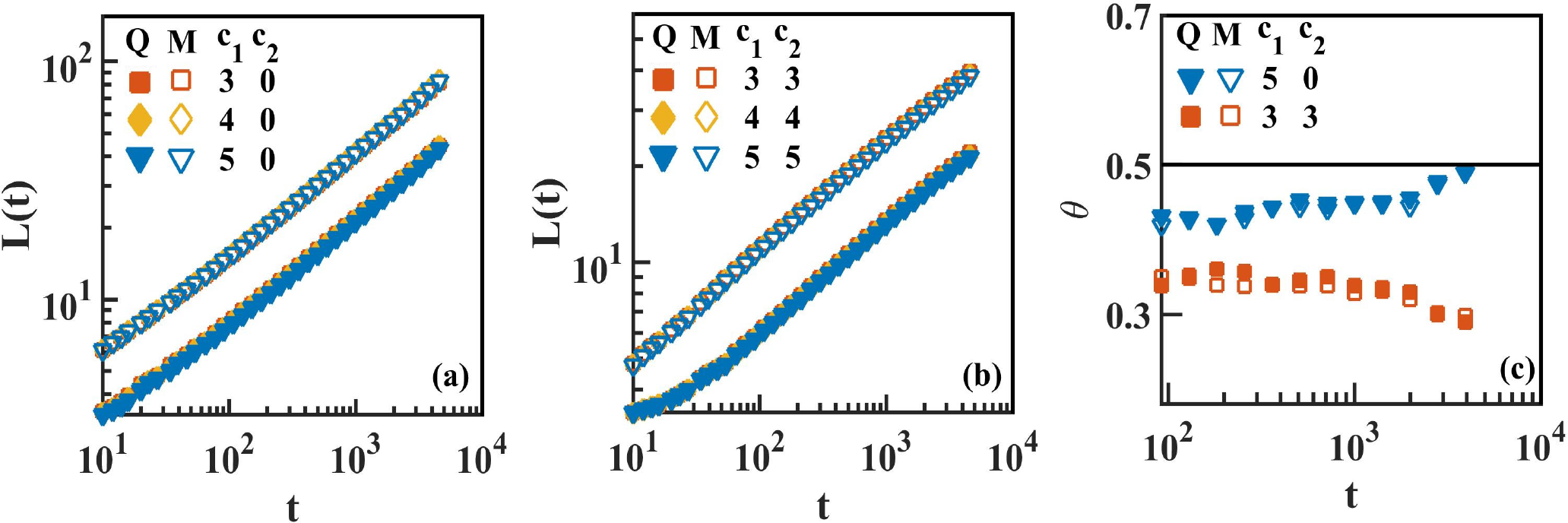} 
\caption{Log-log plot of growth laws in $d=2$ for the nematic (solid symbols) and magnetic (open symbols) components for (a) Case 1(\romannumeral 1), and (b) Case 1(\romannumeral 3). (c) Effective growth exponent $\theta= d(\ln L)/d(\ln t)$ vs. $t$ on a log-linear scale. The solid line corresponds to $\theta = 1/2$. {  The symbol sizes are larger than the error bars.}}  
\label{f4}
\end{figure}

We now study the growth laws in this coupled system. Fig.~\ref{f4} shows length scale data for (a) Case 1(\romannumeral 1) with $c_1=3,\ 4,\ 5$ and $c_2=0$ and (b) Case 1(\romannumeral 3) with $c_1=c_2=c$ for $c=3,\ 4,\ 5$. In both figures, the nematic and magnetic data is depicted by solid and open symbols respectively. Recall that $T_c^N<T<T_c^M$ for this case, but the ordering magnetic component drives the isotropic nematic component via the magneto-nematic coupling. Our reference point is the uncoupled limit, where both fields are characterized by a growth law slower than $t^{1/2}$ due to logarithmic corrections for $d=2$, $n=2$: $L(t) \sim (t/\ln t)^{1/2}$ \cite{Puri_2009,Bray_2002}. Figs.~\ref{f4}(a)-(b) provide the log-log plot of $L(t)$ vs. $t$. Data for both fields exhibit a power-law-like behavior with $L(t)\sim t^{\theta}$. Notice that $L_Q<L_M$ for all times as the ${\bf Q}$-field is slaved to the ${\bf M}$-field. Further, in  Figs.~\ref{f4}(a)-(b), the length scales do not show a significant dependence on the coupling strength. To determine the slopes accurately, we evaluate the effective growth exponent $\theta = d \ln L(t)/d \ln t$. Fig.~\ref{f4}(c) shows $\theta$ vs. $t$ on a semi-log scale. {  This is computed as a discrete derivative, and therefore these plots are known to be quite noisy. In any case, we do not expect a flat behavior as the exponent is not constant. Rather, it is a power-law approximation to the growth law $L(t) \sim (t/\ln t)^{1/2}$.} The solid line denotes $\theta=1/2$. As expected, $\theta$ is effectively reduced due to the logarithmic corrections. For the case $c_1=c_2=c$, we see that $\theta \simeq 0.35$.
\begin{figure}[t]
\centering                      
\includegraphics[width=0.8\linewidth]{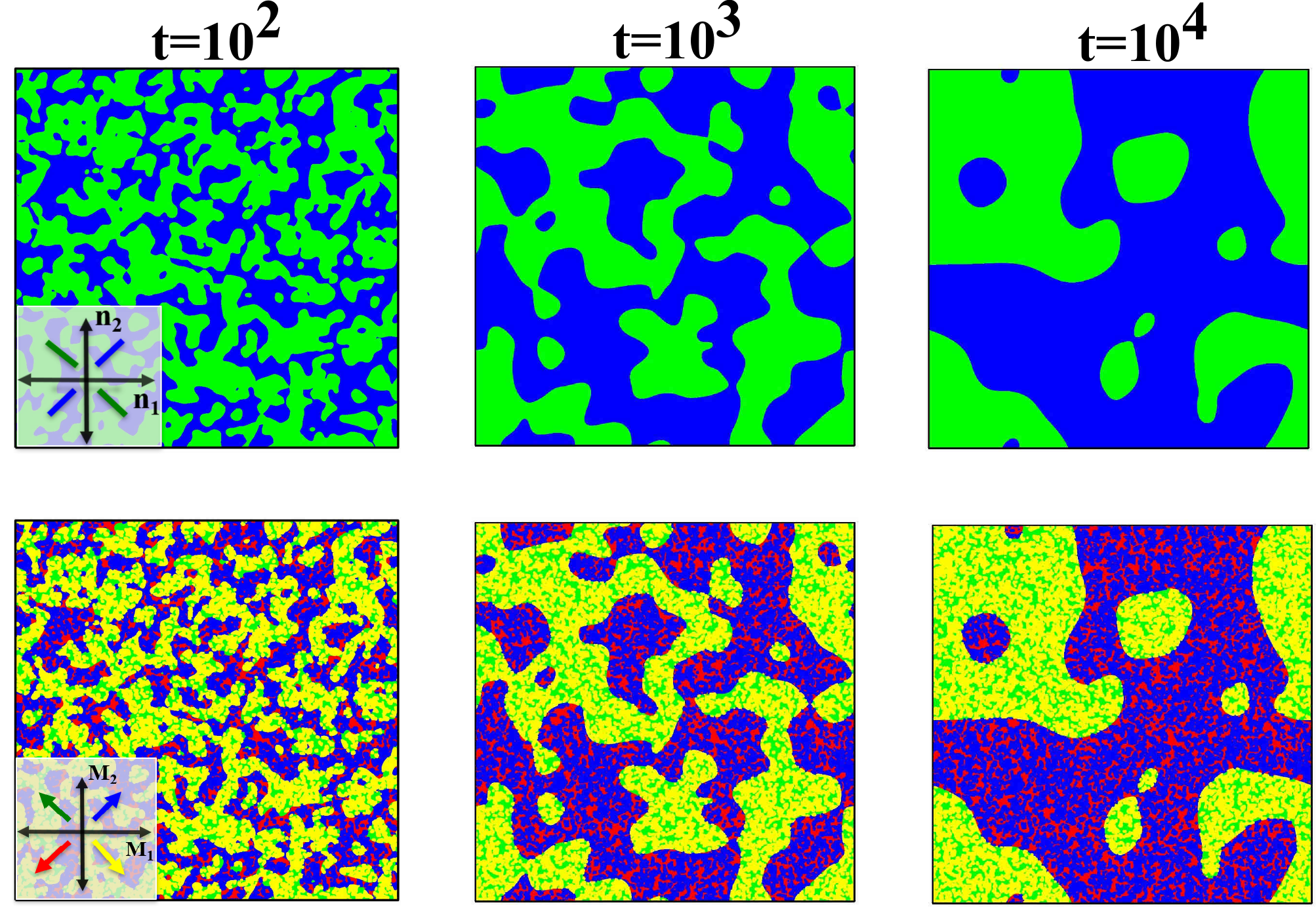} 
\caption{Evolution morphologies in $d=2$ for Case 2 (\romannumeral 2), i.e., $c_1 = 0,\ c_2=4$. We show snapshots at $t=10^2,10^3,10^4$ for the nematic (top) and magnetic (bottom) components.}
\label{f5}
\end{figure}

Next, we study the situation where the magnetic component is influenced by the freely evolving nematic, i.e., $c_1=0, \ c_2\ne 0$.  We focus on Case 2(\romannumeral 2) which 
\begin{figure}[t]
\centering                      
\includegraphics[width=0.9\linewidth]{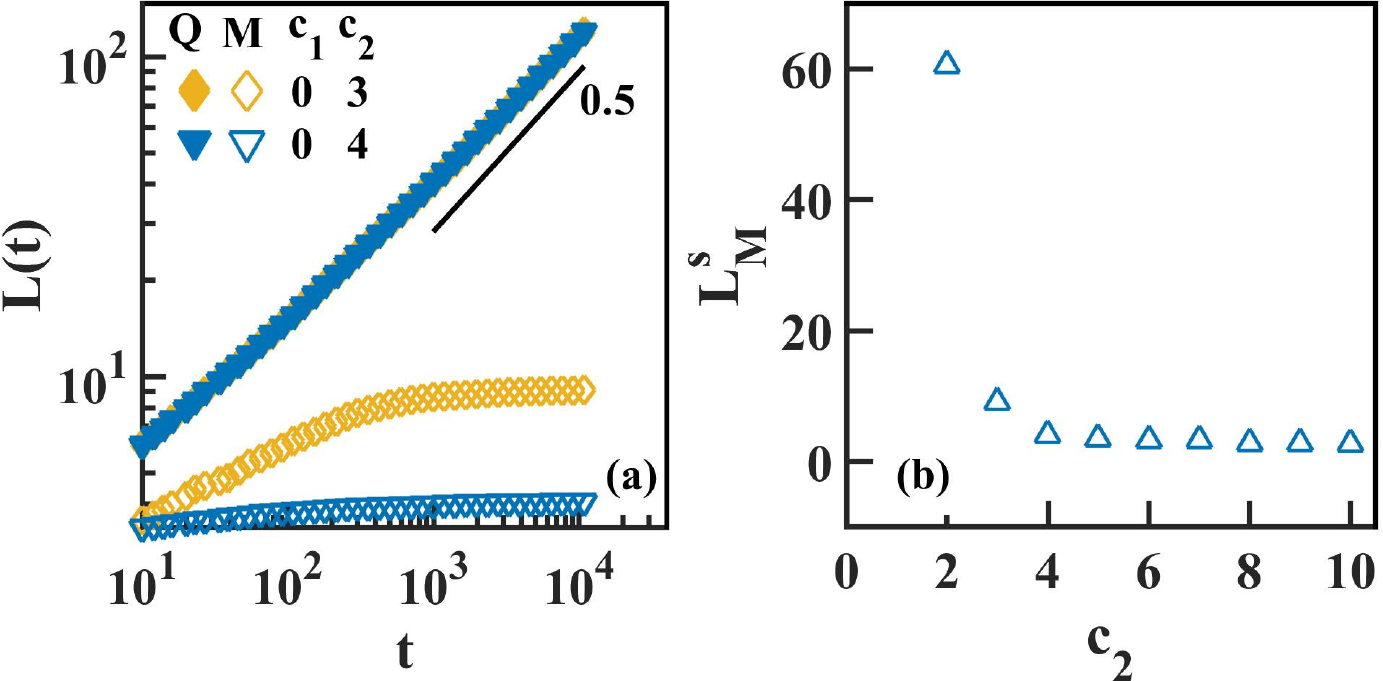} 
\caption{Domain growth in $d=2$ for Case 2(ii). (a) Growth laws for the nematic (solid symbols) and magnetic (open symbols) components. (b) Saturation length $L_M^s$ for different values of $c_2$.}
\label{f6}
\end{figure}
corresponds to $T_c^M<T<T_c^N$. In this case, the linearized calculation shows that ${\bf Q}$ is unstable and ${\bf M}$ is stable about ${\bf Q^*} = 0$, ${\bf M^*}=0$ (the disordered state). The stationary solutions have been explicitly evaluated for this case in Sec.~\ref{s22}, see Eq.~(\ref{SS_case22}). This shows the presence of non-trivial solutions for ${\bf M}$  when $c_2>1$. In Fig.~\ref{f5},  we show nematic (top) and magnetic (bottom) morphologies for $c_1=0,\ c_2=4$ at $t=10^2,\ 10^3,\ 10^4$. Here, the ${\bf M}$-field is linearly stable, so the ${\bf Q}$-field must grow to a certain level before it can enslave the ${\bf M}$-field to grow. The snapshots reveal a sub-domain morphology (SDM) in the magnetic component due to the two possible orientations, ${\bf n}\parallel {\bf M}$ and ${\bf n}\parallel -{\bf M}$. There is a cost to these structures due to surface tension, but the associated entropic gain is large. Further, the magnitudes of {\bf Q} and {\bf M} agree with Eq.~(\ref{SS_case22}). Interestingly, such SDMs have been observed experimentally by Mertelj et al. in FNs \cite{Mert_Na2013}. They obtained a similar poly-domain structure with two opposing states of magnetization, parallel or anti-parallel to the director field, when
\begin{figure}[t]
\centering                      
\includegraphics[width=0.9\linewidth]{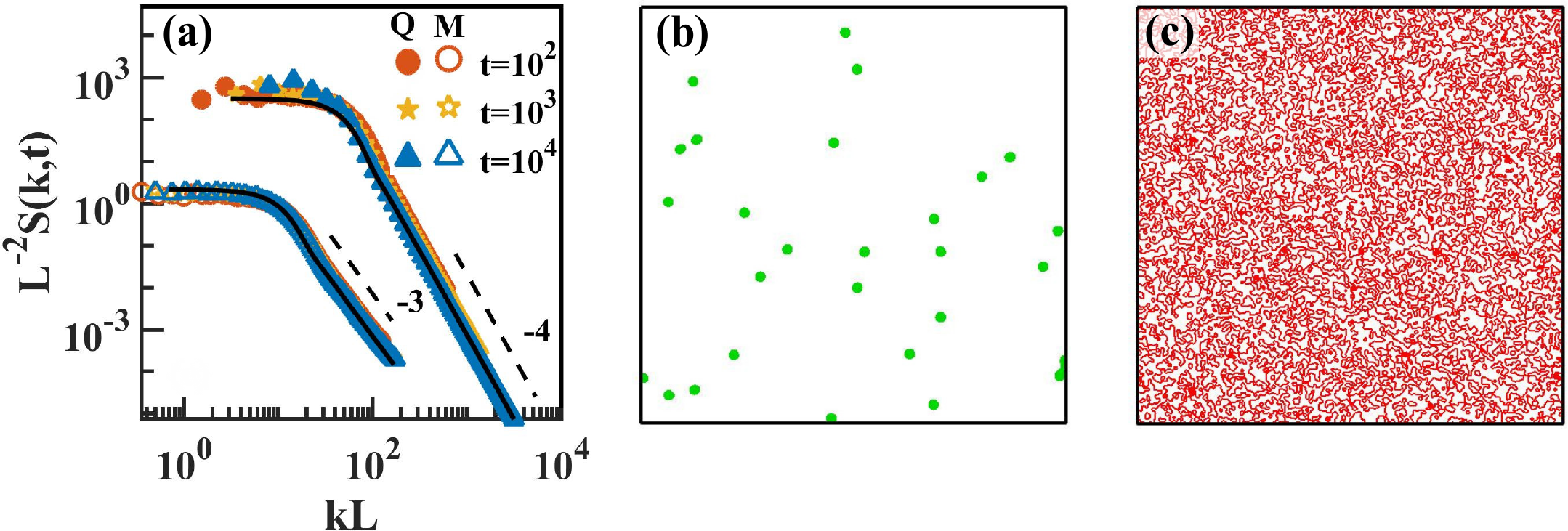} 
\caption{(a) Scaled structure factors for nematic (solid symbols) and magnetic (open symbols) components for Case 2(ii) with $c_1=0,\ c_2=4$. The data sets have been shifted for clarity. The solid lines in (a) denote the BPT function calculated for $n=1,\ d=2$ (M) and $n=2,\ d=2$ (Q). Defect locations of the (b) {\bf Q}-field and (c) {\bf M}-field for morphologies at $t=10^3$ in Fig.~5. The defects are defined as regions where $S<0.6$ and $|{\bf M}|<0.6$, respectively.}
\label{f7}
\end{figure}
the system was quenched in the absence of an external field. In Fig.~\ref{f6}(a), we show the corresponding growth laws for $c_2=3,\ 4$. Notice that $L_Q(t) \sim t^{\theta}$, with $\theta < 0.5$ due to logarithmic corrections.  On the other hand,$L_M(t)$ saturates due to the formation of the SDM. Fig.~\ref{f6}(b) depicts the saturation length scale $L_M^S$ as a function of coupling strength $c_2$ for a $(2048)^2$ system. We do not have a quantitative argument for the dependence of $L_M^S$ on $c_2$, but expect $L_M^S\rightarrow 0$ as $c_2\rightarrow \infty$ and $L_M^S\rightarrow \infty$ as $c_2\rightarrow 1$. 

More insights into the SDM are provided by the scaled structure factor $L^{-2} S(k,t)$ vs. $kL$, plotted in Fig.~\ref{f7}(a), for $c_1=0,\ c_2=4$ at $t=10^2,\ 10^3,\ 10^4$. The data collapses neatly for the ${\bf Q}$ and ${\bf M}$ fields, demonstrating dynamical scaling. The solid lines here are the Fourier transform of BPT function calculated by setting $n=2,\ d=2$ and $n=1,\ d=2$ for nematic and magnetic components respectively. Further, $S_Q(k,t) \sim k^{-4}$ for large $k$, indicating a generalized Porod tail that is dominated by {\it vortex} defects in the nematic morphology. However, $S_M(k,t) \sim {k}^{-3}$, characteristic of scattering from {\it sharp interfaces}. Though ${\bf M}$ is a continuous order parameter, the scattering is dominated by sharp interfaces between ${\bf M}^*$ and  $-{\bf M}^*$ in the SDM. To verify these observations, we show in Figs.~\ref{f7}(b) and (c) the defect locations in the nematic and magnetic fields for the morphologies at $t=10^3$ in Fig.~\ref{f5}. The defects are defined as regions where $S<0.6$ and $|{\bf M}|<0.6$, respectively. Figs.~\ref{f7}(b)-(c) clearly show the dominance of isolated vortex defects in the nematic field and interfacial defects in the magnetic order parameter field. These observations are generic to all the cases in the limit $c_1=0,\ c_2 \ne 0$, see Tables \ref{TabS1}-\ref{TabS3} in Appendix \ref{app1}. An important message here is that the typical length scale in the SDM can be controlled by the coupling strength. We believe this could have experimental implications  for tailoring patterns in FNs.

\subsection{Domain Growth in $d=3$ Ferronematics}
\label{s32}

{  In $d=3$, the ${\bf Q}$-tensor is a symmetric traceless $3\times3$ matrix and Tr$({\bf Q})^3\ne 0$. Consequently, there is a cubic term in the GL free energy [cf. Eq.~(\ref{F_e})]. Then the uncoupled system exhibits a first-order nematic-isotropic transition, unlike the continuous transition in $d = 2$. The $d = 3$ framework also allows for biaxiality which yields richer defect cores, e.g., strings and hedgehogs.}
The ${\bf Q}$-tensor has five independent variables ($\{q_i\}$, $i=1,2,\cdots,5$), and the magnetization ${\bf M}$ has three components ($\{M_j\}$, $j=1,2,3$). The systemic evolution of these eight variables is governed by the eight coupled Eqs.~(\ref{3d_tdgl_1})-(\ref{3d_tdgl_8}) in dimensionless form. {  The mesh sizes are taken to be $\Delta x=1$ and $\Delta t = 0.02$, and are consistent with the requirements of stability.} The equations provide the $q_i$'s and $M_j$'s at each lattice point. The corresponding {\bf Q}-tensor is symmetric and traceless, but not necessarily diagonal. So we choose a reference system aligned with the principal axis that diagonalizes the ${\bf Q}$-tensor. The largest eigenvalue provides $S$ in Eq.~(\ref{QT}), and the corresponding eigenvector is ${\bf n}$, which is needed to characterize the evolution morphology.
\begin{figure}[t]
\centering                      
\includegraphics[width=0.8\linewidth]{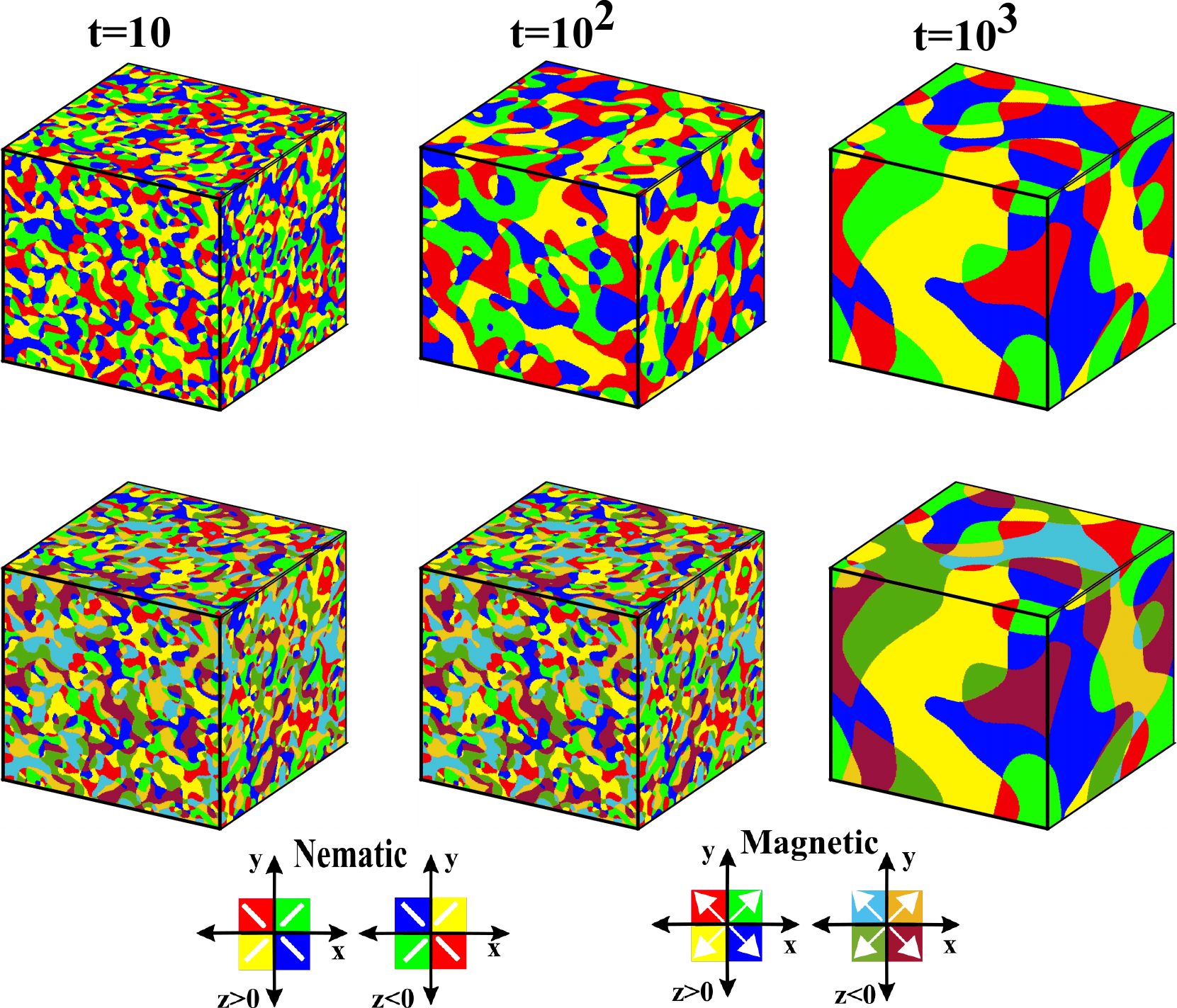} 
\caption{Domain growth for $d=3$ FNs. Nematic (top row) and magnetic (bottom row) morphologies for Case 1(\romannumeral 1) with $c_1=3,\ c_2=0$. We show snapshots at $t=10,\ 10^2,\ 10^3$. The color code used for the snapshots is shown at the bottom.}
\label{f8}
\end{figure} 
In Fig.~\ref{f8}, we depict the coarsening morphologies of the nematic (top row) and magnetic component (second row) for Case 1(\romannumeral 1) in Table \ref{Tab1} with $c_1=3,\ c_2=0$.  Here, the ${\bf Q}$-field undergoes slaved coarsening due to the ${\bf M}$-field. The color scheme for Fig.~\ref{f8} is presented in the bottom row. The ${\bf n}$-field is represented by 4 colors due to the up-down symmetry. The ${\bf M}$-field lacks the up-down symmetry, and is represented by 8 colors. We emphasize here that ${\bf n}$ is parallel to ${\bf M}$ or $-{\bf M}$ due to the magneto-nematic coupling despite the fact that $T_c^N<T<T_c^M$.

We now characterize the $d=3$ morphologies shown in Fig.~\ref{f8} by evaluating
\begin{figure}[t]
\centering                      
\includegraphics[width=0.75\linewidth]{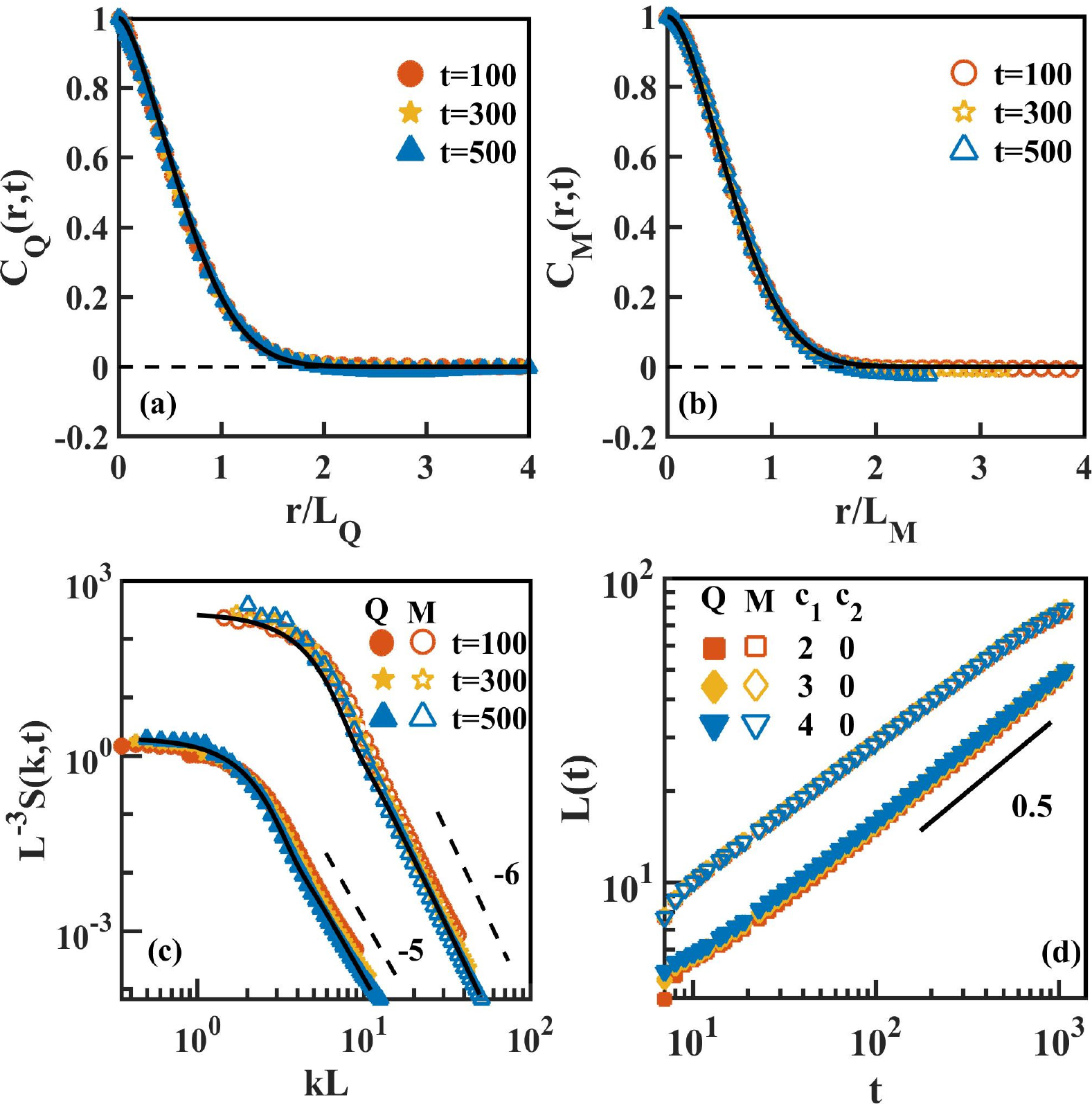} 
\caption{Dynamical scaling of correlation function for the evolution in Fig.~\ref{f8}: (a) nematic component, and (b) magnetic component. The BPT function (solid line) in (a) is evaluated for $n=2,\ d=3$, and in (b) for $n=3,\ d=3$. (c) Scaled structure factors for the nematic (solid symbols) and magnetic (open symbols) components. The corresponding BPT functions are shown as solid lines. The data sets have been shifted for clarity. (d) Growth laws for $c_2=0$ and $c_1=2,3,4$.}
\label{f9}
\end{figure}
the correlation function and structure factor. $C_Q({r},t)$ is calculated directly from the definition in Eq.~(\ref{cr_def}) with ${\psi}\rightarrow P_2(\cos{\theta})$, and $C_M({r},t)$ is calculated as before. Fig.~\ref{f9}(a) shows $C_{Q}({r},t)$ vs. ${r}/L(t)$ for the coupling strength $c_1=3,\ c_2=0$ at $t=100,\ 300,\ 500$ for the ${\bf Q}$-field (solid symbols) and Fig.~\ref{f9}(b) shows the corresponding data for the ${\bf M}$-field (open symbols). The excellent data collapse in both figures demonstrates dynamical scaling for the evolution. $C_Q(r,t)$ data fit well to the BPT function calculated for $n=2$ in contrast to the data for $C_M(r,t)$ which is described 
\begin{figure}[t]
\centering                      
\includegraphics[width=0.7\linewidth]{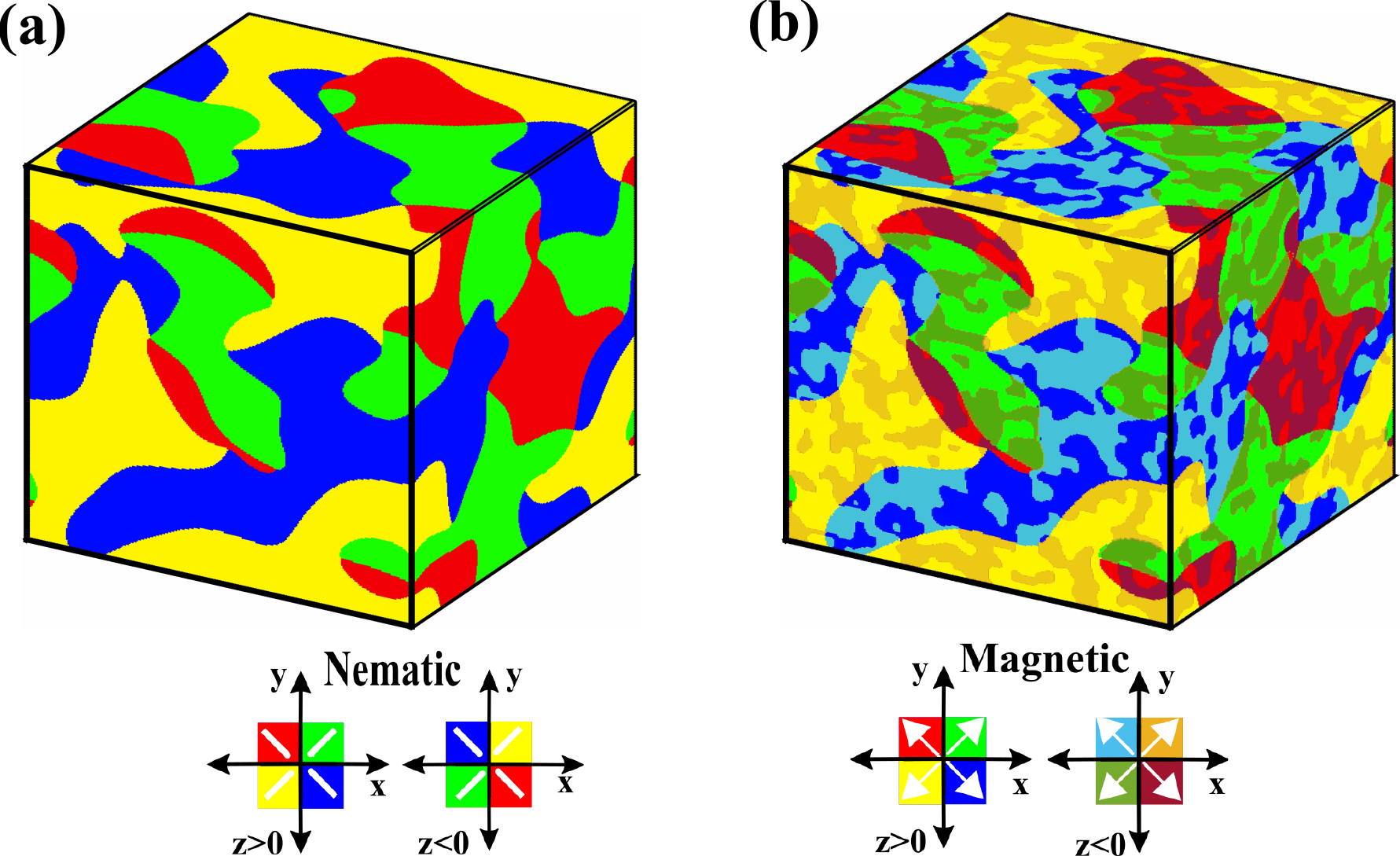} 
\caption{(a) Nematic and (b) magnetic  snapshots in $d=3$ for Case 2(\romannumeral 2), i.e., $c_1=0,\ c_2 = 3$. The snapshots correspond to $t=500$.}
\label{f10}
\end{figure}
well by the BPT function with $n=3$. This clearly demonstrates a distinct behavior at small $r/L$ and results in different defect structures in both components. This will be explained shortly. The corresponding structure factor, $L^{-3}S({k},t)$ vs. ${k}L$, is shown in Fig.~\ref{f9}(c) - the data have been shifted for clarity. As expected, we observe the generalized Porod law $S_M({k},t) \sim {k}^{-6}$ $(d=3,n=3)$ for the ${\bf M}$-field characterizing scattering off monopole defects. However, $S_Q({k},t) \sim {k}^{-5}$, indicating that the slaved nematic morphologies are dominated by $+1/2$ string defects. The difference in the large-$k$ behavior of $S_Q$ and $S_M$ arises from the discrepancy in the small-$r$ behavior in Figs.~\ref{f9}(a)-(b). In Fig.~\ref{f9}(d), we plot $L(t)$ vs. $t$ for both the components on a log-log scale $c_1=2,\ 3,\ 4$. Notice that the length scale for the ${\bf Q}$-field is always less than that for the ${\bf M}$-field. This is a consequence of the lag due to enslavement. The data fits well to the LAC law, i.e., $L(t)\sim t^{1/2}$. There are no logarithmic corrections to the growth law in $d=3$. We have also studied all the other cases; our observations have been summarized in Table \ref{Tab_NR_d=3}  of Appendix \ref{app3}. 
\begin{figure}[t]
\centering                      
\includegraphics[width=0.7\linewidth]{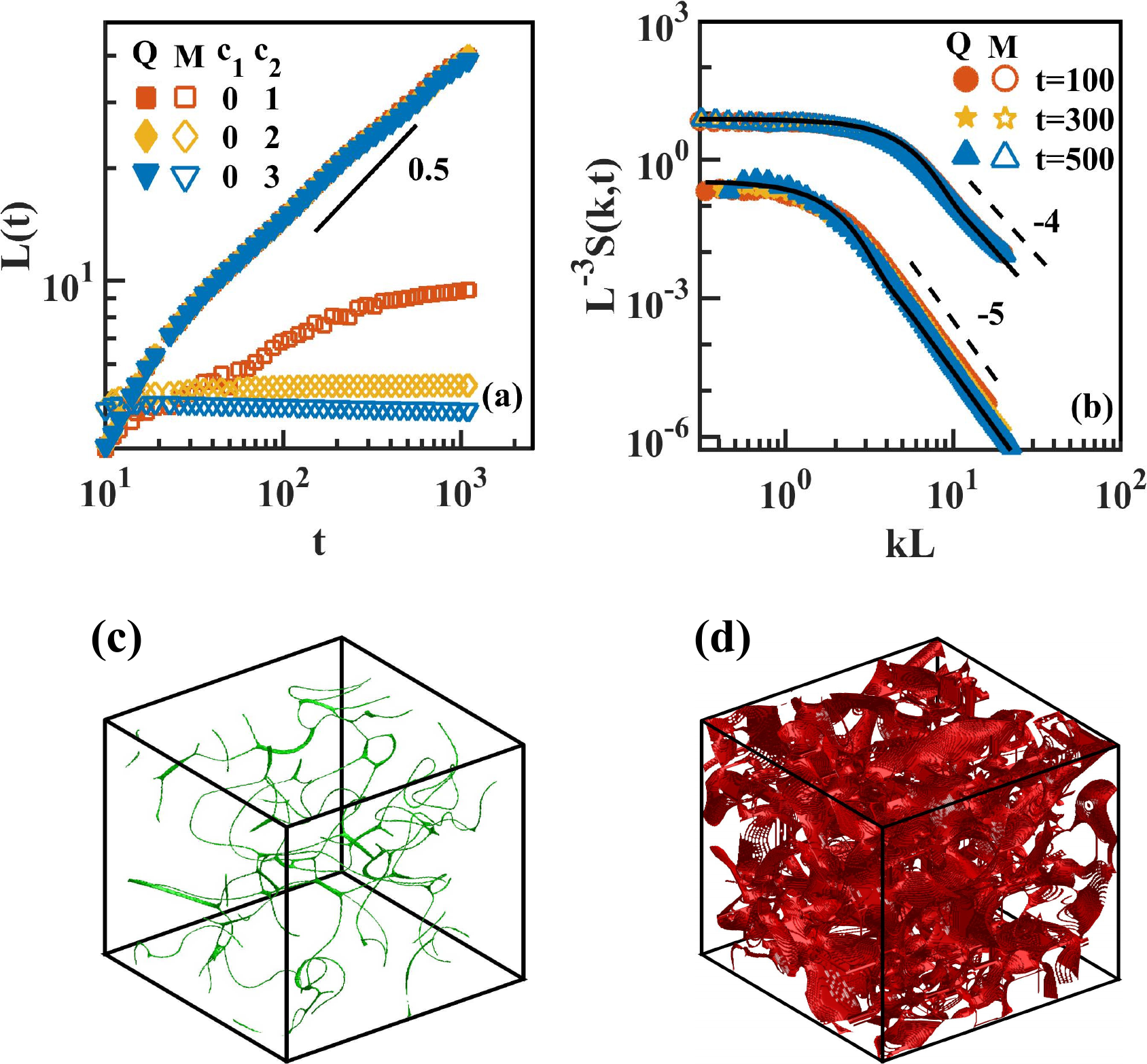} 
\caption{(a) Growth laws for nematic (solid symbols) and magnetic (open symbols) components for Case 2(\romannumeral 2). (b) Dynamical scaling of the structure factor for  $c_1=0,\ c_2=3$. The data has been shifted for clarity. The solid lines denote the BPT function calculated for $n=2,\ d=3$ (Q) and $n=1,\ d=3$ (M). Defect regions in (c) nematic component and (d) magnetic component. These are identified as regions where $S<0.6$ and $|{\bf M}|<0.6$, and are plotted at $t=500$.}
\label{f11}
\end{figure}

Finally, we present some more exotic morphologies which emerge for Case 2 (\romannumeral 2), where the uncoupled system is in the nematic-paramagnetic phase. (We have studied the corresponding $d=2$ case in Figs. \ref{f5}-\ref{f6}.) Figs.~\ref{f10}(a)-(b) show the nematic and magnetic snapshots for $c_1=0,\ c_2=3$ at $t=500$. The co-alignment of ${\bf n}$ and ${\bf M}$ leads to slaved magnetic ordering, but the emergent morphologies show SDM (see Fig.~\ref{f5} for analogous structures in $d=2$ FNs). Fig.~\ref{f11}(a) shows the growth law (left) for the ${\bf Q}$ and ${\bf M}$ components (solid and open symbols) for Case 2 (\romannumeral 2). As expected, $L_Q(t) \sim t^{1/2}$, but $L_M(t) \rightarrow L_M^s$ as $t \rightarrow \infty$ due to the formation of SDM. In Fig.~\ref{f11}(b), we show scaled structure factor plots for $c_1=0,\ c_2=3$. Both data sets exhibit dynamical scaling, confirming the presence of a unique length scale. The  nematic component exhibits the generalized Porod tail, $S_Q(k,t) \sim k^{-5}$, signifying a dominance of string defects $(d=3,n=2)$. On the other hand, we see the usual Porod law [$S_M(k,t)\sim k^{-4}$] for the magnetic component due to scattering off sharp interfaces $(d=3,n=1)$ between the sub-domains with magnetization {\bf M} and -{\bf M}. The excellent fits to the BPT functions calculated for $n=2$ and $n=1$ confirm the different defect structures for nematic and magnetic components respectively. In Figs.~\ref{f11}(c)-(d), we have plotted the defect regions in the nematic and magnetic order parameter fields at $t=500$ for $c_1=0, c_2=3$. The dominance of string defects in the nematic component, and interfacial defects in the magnetic component, are in accordance with observations of structure factor tails in the respective cases.

\section{Summary and Discussion}
\label{s4} 

Let us conclude this paper with a summary and discussion of our results.  In the 1970's, Brochard and de Gennes in pioneering work \cite{Broc_1970} suggested that the possibility of {\it ferronematics} (FNs), formed by the addition of magnetic nanoparticles (MNPs) to nematic liquid crystals (NLCs). They argued that FNs can exhibit spontaneous magnetization, i.e., without any external fields. The first stable suspension was obtained in a seminal experiment by Mertelj et al. \cite{Mert_Na2013}, nearly four decades after it's theoretical conceptualization. Ever since, FNs  have provided new opportunities to study magneto-mechanical and magneto-optic effects in NLCs for diverse applications, ranging from photonics to optical switches, and microfluidics to cosmology. Theoretical studies of FNs are somewhat limited and ours is the first study to discuss the important non-equilibrium phenomenon of coarsening or domain growth. 

FNs are described by two order parameters: the ${\bf Q}$-tensor which describes the nematic order, and the magnetization ${\bf M}$ which describes the magnetic order arising due to the nano-inclusion. We use Ginzburg-Landau-de Gennes free energies with a dyadic magneto-nematic coupling term to capture the surface anchoring between the MNPs and the NLC molecules. We perform temperature quenches, and study the kinetics of phase transitions using coupled time-dependent Ginzburg-Landau (TDGL) equations for ${\bf Q}$ and ${\bf M}$. Our framework has two phenomenological parameters: $c_1$ quantifying the coupling strength of ${\bf Q}$ with ${\bf M}$, and $c_2$ which is the coupling strength of ${\bf M}$ with ${\bf Q}$. In our dimensionless formulation, $c_1$ and $c_2$ arise from the ${\bf Q}-{\bf M}$ coupling term in conjunction with natural order parameter scales. Both the constants originate from the magneto-nematic coupling term in the free energy, and in principle, can be evaluated from experimentally measurable quantities. Their interplay leads to exotic stable nematic and magnetization morphologies, which are not accessible in uncoupled systems. The coupling also leads to the formation of domain walls and point defects, even without the application of external fields. 

The main results of our study are summarized below:\\ 
(\romannumeral 1) For shallow quenches $T_c^N<T<T_c^M$, where $N$ and $M$ refer to nematic and magnetic, the naturally {\it isotropic} nematic component coarsens along with the {\it ordering} magnetic component due to the magneto-nematic coupling. There is formation of nematic and magnetic domains, which are co-aligned. Similar statements hold for shallow quenches such that $T_c^M<T<T_c^N$. {\it Slaved} coarsening is observed in all cases, provided the disordered component is strongly coupled with the ordering component.   \\
(\romannumeral 2) In all cases, we observe the usual Lifshitz-Allen-Cahn (LAC)  law: $L(t)\sim \left(t/\ln t\right)^{1/2}$, in $d=2$, and $L(t) \sim t^{1/2}$ in $d=3$. The corresponding structure factor for both components exhibits {\it generalized Porod decay}: $S(k)\sim k^{-(d+n)}$, where $d$ is the dimensionality. The value of $n$ depends on the nature of the defects in the order parameter field, which can be vortices, strings or monopoles.\\
(\romannumeral 3) For $T_c^M<T<T_c^N$, the asymmetric coupling $c_1=0,\ c_2 \ne 0$ leads to an interesting sub domain morphology (SDM) for the enslaved magnetic component. This arises due to the competition between surface tension and entropic effects, and consists of a microstructure with domains of ${\bf M}^*$ and $-{\bf M}^*$. The typical length scale $L_M^s$ of the  microstructure can be tailored. The structure factor for SDMs exhibits the usual  {\it Porod decay}, $S(k)\sim k^{-(d+1)}$. The latter is characteristic of scattering from sharp interfaces, in contrast to our expectation of the generalized Porod law for scattering from vortex defects in continuous-spin models \cite{Bray_Puri_1991,Toyoki_1992}.

The above observations are experimentally realizable, and our phenomenological set-up can be readily generalized to other systems with are described by coupled order parameters. The study of FNs is an emerging area, rich in fundamental physics and technological applications. We hope that our work initiates joint experimental and theoretical work on this relatively new  soft matter system, and takes it further for futuristic applications, the most significant being the multi-billion dollar liquid crystal display (LCD) industry.

\begin{acknowledgments}
AV acknowledges UGC, India for a research fellowship. AV and VB gratefully acknowledge partial financial support from DST-UKIERI and the HPC facility of IIT Delhi for the computational resources. AV and VB also acknowledge Professor Apala Majumdar from the University of Strathclyde and Konark Bisht from IIT Delhi for illuminating discussions, especially during the formulation of the problem. 
\end{acknowledgments}

\bibliographystyle{apsrev4-1}
\bibliography{slavef} 

\newpage

\appendix
\setcounter{table}{0}
\setcounter{equation}{0}
\renewcommand{\thetable}{A.\arabic{table}}
\renewcommand{\theequation}{A.\arabic{equation}}

\section{Analytical and Numerical Results for Domain Growth in $d=$2}
\label{app1}

In this Appendix, we present analytical results for stable fixed points of our TDGL model for FNs in $d=2$. We consider all the cases in Table \ref{Tab1}. We also summarize numerical results for growth laws and structure factor tails in domain growth.

\begin{center}
\begin{table}[h]
     \centering
      \begin{tabular}{|c|c|c|c|}
            \hline
                \textbf{Coupling limits}& \makecell{ \textbf{Stable stationary solutions}\\($Q_{11}^*,Q_{12}^*,M_1^*,M_2^*$)}&\makecell{\textbf{Growth laws}\\ $\left[L_Q,L_M\right]$}&\makecell{\textbf{Structure factor tails}\\ ($S_Q,S_M$)}\\ 
            \hline
                (\romannumeral 1)\ $c_1\ne0$,\ $ c_2=0$ &  \makecell{($r_Q,0,1,0$) \\ $r_Q=c_1(1+\hat{S})^{-1}$\\ $\hat{S}=S^2/4={3}^{-1}(-2+a_1+{a_1}^{-1})$\\ $a_1={2^{1/3}}(a_2+c_1\sqrt{54+27a_2})^{-1/3}$\\ $a_2=2+c_1^2$\\ } & $\left[(t/\ln t)^{1/2},(t/\ln t)^{1/2}\right]$  &$(k^{-4},k^{-4})$\\
           \hline
               (\romannumeral 2)\ $c_1= 0$,\ $ c_2\ne0$ & $(0,0,1,0)$& $\left[\mbox{No growth},(t/\ln t)^{1/2}\right]$  &$(\mbox{Uniform},k^{-4})$\\
           \hline
              (\romannumeral 3)\ $c_1=c_2=c$\hspace*{.2cm}&\makecell{($r_Q,0,r_M,0$)\\
              $r_Q= c\ (1+\hat{S}+ c^2)^{-1}$\\
              $\hat{S}=S^2/4=(1+c^2)(a_1-2/3)+{a_1}^{-1}$\\
              $3a_1={2^{1/3}}{(33c^2+a_2+{a_3}^{1/2})^{-1/3}}$\\
              $a_3=1053c^4+54c^2a_2$\\
              $a_2=2c^2+6c^4+2c^6$\\  $r_M=\big[(1+\hat{S})(1+c^2+\hat{S})^{-1}]^{1/2}$} & $\left[(t/\ln t)^{1/2},(t/\ln t)^{1/2}\right]$& $(k^{-4},k^{-4})$ \\
       \hline
        \end{tabular}
     \caption{Coupling limits, stable stationary solutions, growth laws and structure factor tails for Case 1 of Table \ref{Tab1}. This corresponds to a quench at temperature $T$ such that $T_c^{N}<T<T_c^{M}$.}
     \label{TabS1}
\end{table}
\end{center}

\begin{center}
\begin{table}[h]
    \centering
      \begin{tabular}{|c|c|c|c|}
           \hline
                \textbf{Coupling limits}& \makecell{ \textbf{Stable stationary solutions}\\($Q_{11}^*,Q_{12}^*,M_1^*,M_2^*$)}&\makecell{\textbf{Growth laws}\\ $\left[L_Q,L_M\right]$}&\makecell{\textbf{Structure factor tails}\\ ($S_Q,S_M$)}\\ 
            \hline
                (\romannumeral 1)\ $c_1\ne0$,\ $ c_2=0$ & $(1,0,0,0)$& $\left[(t/\ln t)^{1/2}, \mbox{No growth}\right] $  &$(k^{-4},\mbox{Uniform})$\\
           \hline
                 ({\romannumeral 2})\ $c_1=0$,\ $ c_2\ne0$ &  \makecell{($1,0,r_M,0$) \\ $r_M=\sqrt{c_2-1}$\\} &  $\left[(t/\ln t)^{1/2},\mbox{Saturation}\right]$  & $(k^{-4},k^{-3})$\\
           \hline
              (\romannumeral 3)\ $c_1=c_2=c$\hspace*{.2cm}&\makecell{($r_Q,0,r_M,0$)\\
               $r_Q= c\ (1-\hat{S} + c^2)^{-1}$\\
              $\hat{S}=S^2/4=0.5(2+c^2+\sqrt{4c^2+c^4})$\\
              $r_M=\big[(\hat{S}-1)(1+c^2-\hat{S})^{-1}]^{1/2}$\\
              } &  $\left[(t/\ln t)^{1/2},(t/\ln t)^{1/2}\right]$  & $(k^{-4},k^{-4})$   \\
         \hline
     \end{tabular}
    \caption{Coupling limits, stable stationary solutions, growth laws and structure factor tails for Case 2 of Table \ref{Tab1}. This corresponds to a quench at temperature $T$ such that $T_c^{M}<T<T_c^{N}$.}
 \label{TabS2}
\end{table}
\end{center}

\begin{center}
\begin{table}[h]
    \centering
      \begin{tabular}{|c|c|c|c|}
              \hline
                    \textbf{Coupling limits}& \makecell{ \textbf{Stable stationary solutions}\\($Q_{11}^*,Q_{12}^*,M_1^*,M_2^*$)}&\makecell{\textbf{Growth laws}\\ $\left[L_Q,L_M\right]$}&\makecell{\textbf{Structure factor tails}\\ ($S_Q,S_M$)}\\ 
              \hline
    (\romannumeral 1)\ $c_1\ne0$,\ $ c_2=0$ &  \makecell{($r_Q,0,1,0$) \\ $r_Q=-c_1(1-\hat{S})^{-1}$\\ $\hat{S}=S^2/4={1}/{3}(-2+a_1+a_1^{-1})$\\ $a_1={2^{1/3}}{(a_2+c_1\sqrt{-54+27a_2})^{-1/3}}$\\} & $\left[(t/\ln t)^{1/2},(t/\ln t)^{1/2}\right] $  & $(k^{-4},k^{-4})$ \\
              \hline 
                  (\romannumeral 2)\ $c_1=0$,\ $ c_2\ne0$ &  \makecell{($1,0,r_M,0$) \\$r_M=\sqrt{c_2+1}$\\} & $\left(t/\ln t\right)^{1/2},\mbox{Saturation})$ & $(k^{-4},k^{-3})$ \\
              \hline
                 (\romannumeral 3)\ $c_1=c_2=c$\hspace*{0.2cm}&\makecell{($r_Q,0,r_M,0$)\\  $r_Q= c\ (1-\hat{S} + c^2)^{-1}$\\ $\hat{S}=S^2/4=0.5(2+c^2+\sqrt{4c^2+c^4})$\\ $r_M=\big[(1-\hat{S})(1+c^2-\hat{S})^{-1}]^{1/2}$\\} & $\left[(t/\ln t)^{1/2},(t/\ln t)^{1/2}\right]$ & $(k^{-4},k^{-4})$  \\
         \hline       
         \end{tabular}    
\caption{Coupling limits, stable stationary solutions, growth laws and structure factor tails for Case 3 of Table \ref{Tab1}. This corresponds to a quench at temperature $T$ such that $T<\mbox{min}\{T_c^{M},T_c^{N}\}$.}
       \label{TabS3}
\end{table}
\end{center}

\newpage
\setcounter{table}{0}
\setcounter{equation}{0}
\renewcommand{\thetable}{B.\arabic{table}}
\renewcommand{\theequation}{B.\arabic{equation}}

\section{TDGL Equations for Ferronematics in $d=$3}
\label{app2}

Taking into account the requirements of symmetry and being traceless, the {\bf Q-}tensor in $d=3$ has five independent parameters:
\begin{eqnarray} 
  {\bf Q}&=&
  \begin{pmatrix}
 -q_1+q_2 & q_3 &  q_4   \\
 q_3 &  -q_1-q_2 & q_5   \\
 q_4 &  q_5 & 2q_1  \\
  \end{pmatrix}.
  \label{Q3}
\end{eqnarray}
It is easy to see that
\begin{eqnarray}
\mbox{Tr} ({\bf Q})&=&0, \\ 
\mbox{Tr}({\bf Q}^2)&=&2q^2= 2(3q_1^2+q_2^2+q_3^2+q_4^2+q_5^2), \\ 
\mbox{Tr}({\bf Q}^3) &=& 6q_1^3-6q_1q_2^2-6q_1q_3^2+3q_1q_4^2+3q_1q_5^2+3q_2q_4^2-3q_2q_5^2+6q_3q_4q_5, \\
\mbox{Tr}({\bf Q}^2)^2 &=& 4(3q_1^2+q_2^2+q_3^2+q_4^2+q_5^2)^2.
\end{eqnarray}

The free energy for the system is given by Eq.~(\ref{F_e}). The TDGL equations for FNs can be written using Eqs.~(\ref{TDGL_genral_1})-(\ref{TDGL_genral_2}). They can be re-scaled into a dimensionless form by introducing the re-scaled variables, $\mathbf{Q} = a\mathbf{Q^\prime}$, $\mathbf{M} = b\mathbf{M^\prime}$, $\textbf{r} = \xi\textbf{r}^\prime$,  $t=\tau t^\prime$. The appropriate choice for the scale factors is
\begin{eqnarray}
a &=& \sqrt{|A|/2B}, \nonumber \\
b &=& \sqrt{|\alpha|/\beta}, \nonumber \\
\xi &=& \sqrt{\kappa/|\alpha|}, \nonumber \\
\tau &=& 1/(|\alpha| \Gamma_M).
\end{eqnarray}
Dropping the primes, we obtain the TDGL equations: 
    \begin{eqnarray}
     \frac{1}{\Gamma}\frac{\partial q_1}{\partial t}&=&\pm 3q_1- q^23q_1 -{\bar{C}}(6q_1^2-2q_2^2-2q_3^2+q_4^2+q_5^2)+l\nabla^2 q_1\nonumber\\
     & &+c_1(-M_1^2-M_2^2 + 2 M_3^2), \label{3d_tdgl_1}\\
     \frac{1}{\Gamma}\frac{\partial q_2}{\partial t}&=&\pm q_2-q^2q_2-{\bar{C}}(4q_1q_2+q_4^2-q_5^2)+l\nabla^2 q_2+c_1(M_1^2-M_2^2),\label{3d_tdgl_2} \\
     \frac{1}{\Gamma}\frac{\partial q_3}{\partial t}&=&\pm q_3-q^2q_3 -{\bar{C}}(-4q_1q_3+2q_4q_5)+l\nabla^2 q_3+2c_1M_1M_2 \label{3d_tdgl_3},\\
    \frac{1}{\Gamma}\frac{\partial q_4}{\partial t}&=&\pm q_4-q^2q_4-{\bar{C}}(2q_1q_4+2q_2q_4+2q_3q_5)+l\nabla^2 q_4+2c_1M_1M_3,\label{3d_tdgl_4}\\
        \frac{1}{\Gamma}\frac{\partial q_5}{\partial t}&=&\pm q_5-q^2q_5 -{\bar{C}}(2q_1q_5-2q_2q_5+2q_3q_4)+l\nabla^2 q_5 +2c_1M_2M_3,\label{3d_tdgl_5}
    \end{eqnarray}
    \begin{eqnarray}
      \frac{\partial M_1}{\partial t} &=&\pm  M_1 -|\mathbf{M}|^2M_1 +\nabla^2 M_1 +c_2[(q_2-q_1)M_1 +  q_3 M_2 +  q_4 M_3],\label{3d_tdgl_6}\\
      \frac{\partial M_2}{\partial t} &=&\pm M_2 -|\mathbf{M}|^2M_2 +\nabla^2 M_2 +c_2[-(q_1+q_2)M_2 +  q_3 M_1 +  q_5 M_3],\label{3d_tdgl_7}\\
      \frac{\partial M_3}{\partial t} &=& \pm M_3 - |\mathbf{M}|^2M_3 + \nabla^2 M_3 +c_2[2 q_1 M_3 +  q_4 M_1 + q_5 M_2].\label{3d_tdgl_8}
   \end{eqnarray}
Here, the dimensionless parameters are
\begin{eqnarray}
\Gamma &=& \dfrac{2 |A| \Gamma_Q}{|\alpha| \Gamma_m}, \nonumber \\
{\bar{C}} &=& \dfrac{C}{2\sqrt{2|A|B}}, \nonumber \\
l &=& \dfrac{L|\alpha|}{2\kappa |A|}, \nonumber \\
c_1 &=& \dfrac{\gamma \mu_0 |\alpha|}{4\beta|A|}\sqrt{\dfrac{2B}{|A|}}, \nonumber \\
c_2 &=& \dfrac{\gamma \mu_0}{2|\alpha|}\sqrt{\dfrac{|A|}{2B}}.
\end{eqnarray}
As discussed in the context of the TDGL equations in $d=2$ [Eqs.~(\ref{2TDGL1})-(\ref{2TDGL4})], the $\pm$ sign with the first term of the right-hand-side depends on whether the quench temperature is above ($+$) or below ($-$) the critical temperature of the corresponding component. The factor $\Gamma$ is the scaled diffusion constant which sets the relative time scale of the two components in the coarsening process. In our study, we set $\Gamma=1$,  $C=1$ and $l=1$. These parameters determine the order of the transition. The constants $c_1$ and $c_2$ are the re-scaled coupling parameters.

As in the $d=2$ case, we can obtain the FPs of the dynamics by setting $\partial/\partial t \equiv 0$, $\nabla^2 \equiv 0$ in Eqs. (\ref{3d_tdgl_1})-(\ref{3d_tdgl_8}). The resultant coupled equations are analytically intractable and have to be solved numerically, in general. Of course, there is a trivial FP with ${\bf Q}^*=0$, ${\bf M}^*=0$, corresponding to the disordered state.

\newpage
\setcounter{table}{0}
\setcounter{equation}{0}
\renewcommand{\thetable}{C.\arabic{table}}
\renewcommand{\theequation}{C.\arabic{equation}}

\section{Numerical Results for Domain Growth in $d=3$}
\label{app3}

In this Appendix, we summarize numerical results for domain growth in $d=3$ FNs.

\begin{center}
\begin{table}[h]
    \centering
      \begin{tabular}{|c|c|c|c|}
              \hline
                    \textbf{Quench temperature}& \makecell{\textbf{Coupling limits} }&\makecell{\textbf{Growth laws}\\ ($L_Q,L_M$)}&\makecell{\textbf{Structure factor tails}\\ ($S_Q,S_M$)}\\ 
              \hline
                   \makecell{Case 1\\ ($T_c^N<T<T_c^M$)}  &  \makecell{(\romannumeral 1)\ $c_1\ne0,\ c_2=0$\\(\romannumeral 2)\ $c_1=0,\ c_2\ne0$\\(\romannumeral 3)\ $c_1=c_2=c$~~~~} & \makecell{$(t^{1/2},t^{1/2})$\\$(\text{No growth},t^{1/2})$\\$(t^{1/2},t^{1/2})$}  & \makecell{$(k^{-5},k^{-6})$\\$(\text{Uniform},k^{-6})$\\$(k^{-5},k^{-6})$} \\
              \hline     
                          \makecell{Case 2\\ ($T_c^M<T<T_c^N$)}  &  \makecell{(\romannumeral 1)\ $c_1\ne0,\ c_2=0$\\(\romannumeral 2)\ $c_1=0,\ c_2\ne0$\\(\romannumeral 3)\ $c_1=c_2=c$~~~~} & \makecell{$(t^{1/2},\text{No growth}))$\\$(t^{1/2},\text{Saturation})$\\$(t^{1/2},t^{1/2})$}  & \makecell{$(k^{-5},\text{Uniform})$\\$(k^{-5},k^{-4})$\\$(k^{-5},k^{-6})$} \\
                          \hline
                            \makecell{Case 3\\ ($T<\text{min}\{T_c^N,T_c^M$\})}  &  \makecell{(\romannumeral 1)\ $c_1\ne0,\ c_2=0$\\(\romannumeral 2)\ $c_1=0,\ c_2\ne0$\\(\romannumeral 3)\ $c_1=c_2=c$~~~~} & \makecell{$(t^{1/2},t^{1/2})$\\$(t^{1/2},\text{Saturation})$\\$(t^{1/2},t^{1/2})$}  & \makecell{$(k^{-5},k^{-6})$\\$(k^{-5},k^{-4})$\\$(k^{-5},k^{-6})$}\\
         \hline      
         \end{tabular}
    
\caption{Quench temperature, coupling limits, growth laws and structure factor tails for nematic and magnetic components in $d=3$.}
       \label{Tab_NR_d=3}
    \end{table}

\end{center}

\end{document}